\documentclass[aps,pre,twocolumn,showpacs,superscriptaddress]{revtex4}
\usepackage{amsmath}
\usepackage{amsfonts}
\usepackage{amssymb}
\usepackage{epsfig}
\usepackage{graphicx}
\renewcommand{\phi}{\varphi}
\newcommand{\be}{\begin{equation}}
\newcommand{\ee}{\end{equation}}
\newcommand{\ba}{\begin{align}}
\newcommand{\ea}{\end{align}}
\newcommand{\ave}[1]{\langle {#1} \rangle}
\usepackage{color}

\begin{document}


\title{Dynamic criticality at the jamming transition}

\author{Atsushi Ikeda}
\affiliation{Laboratoire Charles Coulomb, UMR 5221, CNRS and Universit\'e
Montpellier 2, Montpellier, France}

\author{Ludovic Berthier}
\affiliation{Laboratoire Charles Coulomb, UMR 5221, CNRS and Universit\'e
Montpellier 2, Montpellier, France}

\author{Giulio Biroli}
\affiliation{Institut de Physique 
Th\'eorique (IPhT), CEA, and CNRS URA 2306, 91191 
Gif-sur-Yvette, France}

\date{\today}

\begin{abstract}
We characterize vibrational motion occurring at low temperatures
in dense suspensions of soft repulsive spheres over a broad range
of volume fractions encompassing the jamming transition
at ($T=0$, $\phi=\phi_J$). We find that  
characteristic time and length scales of thermal vibrations
obey critical scaling in the vicinity of the 
jamming transition. We show in particular that the amplitude and 
the time scale of dynamic fluctuations diverge symmetrically on both sides 
of the transition, and directly reveal a diverging correlation 
length.
The critical region near $\phi_J$ is divided in three
different regimes separated by a characteristic temperature 
scale $T^\star(\phi)$ that vanishes quadratically with the distance to 
$\phi_J$. 
While two of them, ($T<T^\star(\phi)$, $\phi>\phi_J$)
and ($T<T^\star(\phi)$, $\phi < \phi_J$), are described by harmonic theories 
developed in the zero temperature limit, the third one for   
$T>T^\star(\phi)$ is inherently anharmonic and displays new critical 
properties. We find that the quadratic scaling of $T^\star(\phi)$ 
is due to nonperturbative anharmonic 
contributions, its amplitude being orders of magnitude smaller 
than the perturbative prediction based on the expansion to quartic order
in the interactions.
Our results show that thermal vibrations in colloidal assemblies 
directly reveal the critical nature of the jamming transition. 
The critical region, however, is very narrow and has not yet been 
attained experimentally, even 
in recent specifically-dedicated experiments. 
\end{abstract}


\maketitle

\section{Introduction}

Important theoretical efforts have recently been devoted to the study of the 
jamming transition and the critical properties of the so-called 
J-point~\cite{reviewmodes}. 
An important feature of a system approaching the jamming transition 
is the emergence of a peculiar density of states which differs
from the ones of ordinary solid materials. The studies of athermal packings 
of soft repulsive particles above the jamming 
transition~\cite{wyart2,silbert,xu,vitelli,vitelli2,recentwyart}, and of 
hard particles below jamming~\cite{brito} both
unveiled that the density of states acquires an excess 
of modes at low frequencies, and thus increasingly differs 
from the usual Debye behavior as jamming is approached.
The presence of an excess of harmonic modes which are distinct from 
usual plane waves has been proposed as an explanation to 
many unusual low-temperature 
properties of disordered materials~\cite{reviewmodes}.  

Motivated by these theoretical results, several recent experiments 
aimed at measuring the vibrational motion of a 
variety of dense colloidal 
assemblies~\cite{bonn,bonn2,soft,craig,craig2,xuprl,caswell}.
Experimentally, the normal modes are accessed by diagonalizing 
the displacement correlation matrix to obtain quantitative information
about time scales (eigenvalues) and length scales (eigenvectors)
of the vibrational motion. These experimental studies were made possible 
through the development of increasingly accurate tools to record 
single particle
motion in colloidal assemblies. Similar studies for standard molecular glasses 
are not possible since particle motion cannot be visualized directly in 
that case, but vibrational motion can also be accessed in systems 
such as driven granular media, where attempts to characterize modes were
recently reported~\cite{britosoft}. 

We wish to characterize the relation, if any, 
between these two sets of efforts. Are the normal modes 
determined in experimental work related to the ones studied 
theoretically and numerically near the jamming transition?
What are the necessary conditions to observe experimentally the 
`low-frequency' or `soft' modes discussed theoretically? 
What are the real space signatures in single particle trajectories
of the anomalous density of states that develops near jamming?
 
More fundamentally, these questions raise the issue of the validity 
of a harmonic description of the dynamics of amorphous materials at 
low temperature.
Indeed, theoretical studies are performed in the zero-temperature limit 
where a harmonic description (or an effective one for 
hard spheres~\cite{brito}) is justified. On the other hand experiments
are inevitably performed at finite temperatures in colloidal systems, 
or under gentle vibrations for granular materials. Therefore, 
understanding to what extent finite temperatures affect 
the theoretical predictions and quantifying the role 
of anharmonicity are two important problems which are 
tackled by the present work. 

The fact that finite temperature can be an extremely singular 
perturbation close to jamming 
can easily be grasped recalling that 
an excess of low energy modes may lead to large fluctuations which can
eventually destabilize the system, as it happens for 
two-dimensional crystals~\cite{chaikin}.
 For the jamming problem, there exist
two possibilities: 
(1) The harmonic description remains 
valid at low enough temperature. Then, the flat density of states 
would imply arbitrarily large fluctuations and would suggest 
the melting of the disordered 
solid, which would then invalidate a 
purely harmonic description.
(2) The harmonic description breaks down approaching the 
jamming transition {\it at any value of the temperature}. 
This would allow the existence of a disordered solid but would also imply 
that anharmonicity is crucial to prevent the solid from breaking apart. 
In both cases, the conclusion is that the harmonic description 
must break down at any finite temperature close enough to
$\phi_J$. Because 
experimental~\cite{bonn,soft,craig,xuprl,caswell} and 
numerical~\cite{tom,hugosoft,hugopre,hayakawa} 
works show that close to jamming a disordered solid 
is stable, the second possibility is clearly the correct one, 
and anharmonicity must play an increasing role approaching the 
jamming transition. This issue was recently discussed 
in Refs.~\cite{vitelli,ohern,dauchotmodes}, but our conclusions 
will differ somewhat from these earlier studies. 
 
In this work we provide  
a complete description of the role of finite temperatures
and anharmonicity on the critical behavior close to the jamming transition and 
the understanding of the connection, or the lack thereof, between 
vibrational properties studied in experiments and theory. We explain 
when it is correct to assume that thermal vibrations and eigenmodes of 
the dynamical matrix coincide, and the range of parameters where
single particle motion becomes dominated by genuine 
`soft' or `low-frequency' modes of the type discussed in the 
context of the jamming transition. This issue is also important theoretically,
as it could help us to understand better the validity of analytical
replica calculations where thermal fluctuations 
are taken into account~\cite{PZ,hugo}, but vibrational eigenmodes
are not directly described. 

Finally, because we focus on thermal 
fluctuations, we cannot, in principle, address the more complicated situation
where particle motion does not result from an equilibrium thermal bath, as 
is the case in granular media~\cite{britosoft,dauchotmodes}.   
However, the type of `vibrational heterogeneity' that we reveal
through the study of thermal vibrations is strongly 
reminiscent of the findings reported in two distinct 
experimental studies of dynamic heterogeneity in driven granular media across 
the jamming transition~\cite{leche,corentin}. In particular, we obtain at
finite temperature a nonmonotonic dependence of both a dynamic 
susceptibility and a correlation length scale very similar 
to the experimental findings for dissipative grains. 
They stem, in our case, from 
the existence of a correlation length that is divergent when the 
temperature is strictly zero. Surprisingly, 
this divergent length scale is physically distinct from 
the `isostatic' length scale which is central to 
the description of the density of states
near jamming~\cite{wyart2}, or the response function of athermal
packings~\cite{saarloos}.  We will connect this last finding with 
recent investigations of random networks~\cite{recentwyart}.

This paper is organized as follows.  
In Sec.~\ref{model} we describe our numerical model and methods. 
In Sec.~\ref{dynamic}, we present numerical results for the dynamic
behavior of low-temperature packings of soft repulsive spheres 
across the jamming density. 
In Sec.~\ref{critical}, we analyze more precisely the scaling behavior 
of these dynamic quantities, and the emergence of diverging time scales
and length scales near jamming. 
In Sec.~\ref{lengthscale}, we show that these divergences result from the 
existence of diverging dynamic fluctuations and of a diverging correlation
length scale. 
The finite temperature critical behavior is analyzed in 
Sec.~\ref{harmonic}.
In Sec.~\ref{anharmonic} we numerically 
determine the temperature scale for the emergence 
of anharmonicity, and provide a theoretical analysis of 
its physical content.
Finally, we use our findings to discuss
experimental results in Sec.~\ref{discussion}.

\section{Model and numerical methods}

\label{model}

We use molecular dynamics simulations to study the thermal vibrations of 
sphere packings near the jamming transition. 
We focus on monodisperse harmonic spheres~\cite{durian}, 
interacting through a 
pairwise potential, 
\be
v(r_{ij}) =  \frac{\epsilon}{2} 
(1-r_{ij}/\sigma)^2  \Theta(\sigma - r_{ij}) ,
\ee 
where $\Theta(x)$ is the Heaviside function and 
$\sigma$ is the particle diameter. 
We use system sizes between $N=120$ and $N=64000$ particles 
imposing periodic boundary conditions, changing the volume 
$V$ to adjust the packing fraction $\phi = \pi \sigma^3 N / (6V)$.
The main results 
are reported for $N=8000$, while we use larger and smaller systems
for specific purposes to be specified below. 
We use Newtonian dynamics, and integrate 
Newton's equation of motion in the microcanonical ensemble
after proper thermalization of the system at the desired temperature.
The results are reported in units of $\sqrt{m\sigma^2/\epsilon}$
for time scales, of $\sigma$ for length scales, and of 
$\epsilon/k_B$ for temperatures, where $k_B$ is the Boltzmann
constant.

To produce a low temperature amorphous packing of harmonic spheres 
in three dimensions at volume fraction $\phi$, we first
perform an instantaneous quench from $T = \infty$ to 
$T=10^{-5}$ (well below the glass temperature~\cite{tom})
at a large volume fraction, $\phi=0.70$.
We then let the system relax and rapidly settle 
in a metastable state. At these very low temperatures (compared to the 
glass temperature) aging effects rapidly become negligible
and the system simply performs vibrations near a well-defined  
energy minimum.

We then explore the $(\phi,T)$ 
phase diagram of these packings by decreasing $T$ and $\phi$
to the desired values. Using this method 
we can follow {\it the same jammed configuration} 
with different particle overlap or gap when changing 
temperature or density.
For each independent quench, the jamming density $\phi_J$ is 
therefore well-defined and can be measured very 
accurately~\cite{ohern1,pinaki}, 
for instance by decompressing the packing at $T=0$ towards 
$\phi_J$ and measuring the volume fraction where the 
pressure vanishes. Following previous
work~\cite{hugopre}, we mainly discuss the 
vibrational behavior of a single, large packing 
with $N=8000$ and for which $\phi_J \simeq 0.6466$. We provide additional
data for different packings where necessary. We checked that the results
reported in this work do not depend on the specific choice of 
a packing.

Note that all dynamic observables discussed in this 
paper are measured after aging effects have terminated, and are thus
stationary or independent of the waiting time, 
even though the dynamics is not ergodic.
We merely explore the statistical properties of a representative 
metastable state. Thus, our studies are qualitatively distinct from 
the non-stationary aging dynamics studied in Ref.~\cite{brito},
and from the athermal studies of Refs.~\cite{teitel,claus,ikeda} where 
an externally imposed shear flow affects the structure and dynamics
of the packings.

\section{Dynamical behavior at finite temperatures 
near jamming}

\label{dynamic}

\subsection{Mean-squared displacement}

In order to characterize vibrational motion, we focus on 
single particle dynamics. The most basic correlation function is the
mean-squared displacement (MSD), which is easily measured experimentally
in colloidal systems, 
\be 
\Delta^2_0(t) = 
\frac{1}{N} \sum_{i=1}^{N} \langle |\Delta {\bf r}_i(t)|^2 \rangle,  
\label{delta}
\ee
where $\Delta {\bf r}_i(t) = {\bf r}_i(t) - {\bf r}_i(0)$, and ${\bf r}_i(t)$ 
represents the position of particle $i$ at time $t$. 
The brackets in Eq.~(\ref{delta}) represent an average 
over long trajectories performed at a given $T$ and $\phi$
(but nevertheless restricted to a single metastable state).
We will be particularly interested in 
the long-time limit of the MSD, which we call the Debye-Waller 
(DW) factor.

Near the jamming transition, there is a slight technical 
problem arising in these measurements because the packings 
produced numerically contain a small fraction of `rattlers', that have 
peculiar dynamic properties since they are less connected than most 
particles and therefore have distinct dynamical properties. An annoying 
feature is that rattlers move much more than the other particles,
and so they might completely dominate the sum in Eq.~(\ref{delta}), thus 
yielding results which do not reflect the representative behavior
of the system. 

We used two distinct methods to 
prevent rattlers from spoiling the measurements.
A first, simple method consists in
computing a modified version of the MSD,  
\be
\Delta^2_{\rm mod}(t) = \frac{1}{N} \sum_{i=1}^N \frac{3}{\langle 
|\Delta {\bf r}_i(t)|^{-2} \rangle},
\label{delta1}
\ee
which gives nearly no statistical weight to the particles 
that move much further than the average. 
The prefactor $3$ in the definition (\ref{delta1}) is such that
$\Delta^2_{\rm mod}(t) = \Delta^2_0(t)$ when the distribution 
of single particle displacement is Gaussian.
In a second method, we directly remove the rattlers and measure 
\be
\Delta^2(t) = \frac{1}{N'} \sum_{i=1}^{N'} \langle |\Delta 
{\bf r}_i(t)|^2 \rangle,  
\label{delta2}
\ee
where $N'$ is the total particle number which are not rattlers.
Although rattlers can be defined unambiguously in the jammed regime, 
$\phi>\phi_J$, by measuring contact numbers in the ground state 
at $T=0$, we cannot apply this method to the unjammed regime, $\phi<\phi_J$. 
We have applied the following procedure which can be used in both regimes. 
We first measure the DW factor of each particle. We then calculate the 
median value, which is less affected by rattlers than the averaged value. 
Then we regard the particles whose DW factor is 5 times larger 
than the intermediate value as rattlers. 
We have confirmed that this procedure and the analysis at the 
ground state give the same result in the jammed regime.  
We have also confirmed that the definitions Eqs.~(\ref{delta1}) 
and (\ref{delta2}) yield similar results at all densities 
and temperatures. We note that 
definition (\ref{delta1}) is conceptually extremely simple 
and does not require any empirical criterion to identify rattlers, and could
therefore be very useful for the analysis of experimental data. 
In the following, we present the data obtained from 
the definition (\ref{delta2}).

\begin{figure}
\psfig{file=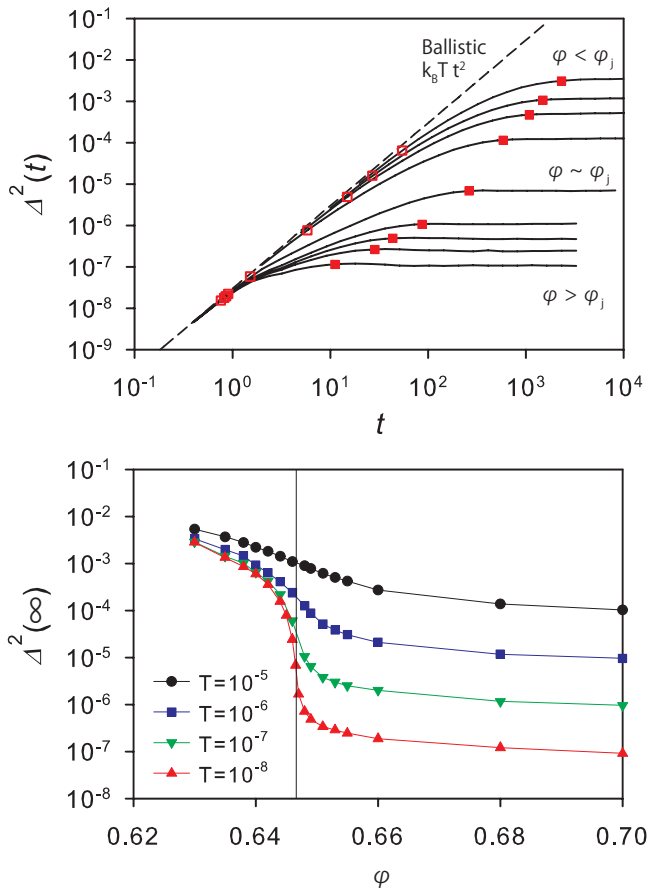,width=8.5cm,clip}
\caption{\label{fig1} 
Top: Time dependence of the 
mean-squared displacements (MSD) at constant temperature, 
$T=10^{-8}$, and various volume fractions increasing from 
top to bottom across the jamming transition.
Open squares indicate the microscopic time scale 
$\tau_0$, filled squares indicate the long timescale $t^\star$,
which marks roughly the convergence of the MSD to its 
long-time limit. Both time scales decrease with $\phi$, but
their ratio is maximum near $\phi_J$. 
Bottom: Volume fraction dependence of the Debye-Waller (DW) factor 
(long-time limit of the MSD) for different temperatures. 
The $T = 0$ jamming singularity at $\phi_J \approx 0.6466$ 
(vertical line) strongly influences the DW factor when 
$T$ becomes very small, $T < 10^{-7}$, and $\phi$ is very close to 
the critical density, $0.64 < \phi_J < 0.655$. }
\end{figure}

In Fig.~\ref{fig1}, we present typical results 
for the behavior of the MSD obtained at constant temperature, $T=10^{-8}$, 
and a range of volume fraction encompassing the jamming density $\phi_J$.
The time dependence of the MSD is typically characterized by 
two distinct time scales. At very short times, the MSD is purely ballistic
and follows $\Delta^2(t) = T t^2$, which simply results from the
short-time integration of Newton's equation of motion for particles
thermalized at temperature $T$. Deviations from this trivial 
behavior occur after a microscopic time scale, $\tau_0$,   
which corresponds to the time where particles start to feel 
the influence of their neighbors. This is indicated by the open squares in 
Fig.~\ref{fig1}. We shall characterize $\tau_0$ more 
accurately in Sec.~\ref{sec:tau0} below, but the data in Fig.~\ref{fig1}
indicate that $\tau_0$ decreases with $\phi$.   

For times $t > \tau_0$, the time evolution of the MSD crosses over to
an intermediate time regime where it obeys diffusive behavior, before
saturating at long times to a plateau, the DW factor $\Delta^2(\infty)$,
which decreases rapidly upon compression.  The approach to the 
plateau value occurs after a typical time scale, $t^\star$, marked
by the filled squares in Fig.~\ref{fig1}. 
The precise definition of $t^\star$ is also discussed below. 
The data indicate again that $t^\star$ 
decreases strongly when the system is compressed. 
The fact that the DW  factors remain finite in the long-time limit
shows that our systems are genuine amorphous solids~\cite{rmp},
and that they do not melt despite having a flat density 
of states at $T=0$ and $\phi = \phi_J$.

We now extract the Debye-Waller factor from the long-time limit
of the mean-squared displacements. We gather our results
for a broad range of densities and temperatures 
in Fig.~\ref{fig1}. We find that the DW factor decreases
upon compression at fixed temperature, mirroring the 
behavior of both time scales $\tau_0$ and $t^\star$. When the system
gets denser, the particles have less space to move, their 
vibrations are more constrained, and so 
the typical amplitude of the vibrations and the related time scales 
become smaller. Therefore, a smooth evolution of dynamic quantities 
with density can be anticipated from trivial reasons, unrelated
to the specific physics of the jamming transition. 

Two interesting features emerge from the DW data in Fig.~\ref{fig1}.
Firstly, the scaling with temperature is different on both sides 
of the jamming transition. This is not surprising, as 
many other observables have a similar dual behavior~\cite{hugo},
which corresponds to the two different situations obtained when 
$T \to 0$. While the system corresponds to a hard sphere assembly
when $\phi < \phi_J$, it is instead a jammed soft solid 
for $\phi > \phi_J$. Accordingly, the DW factor 
converges to its hard sphere value below jamming 
and remains finite as $T \to 0$, 
while it vanishes linearly with $T$ above jamming, as in ordinary 
solids~\cite{chaikin}.
 
Secondly, we also note that as $T$ becomes smaller, the 
volume fraction dependence of the DW factor becomes more and 
more singular in the vicinity of $\phi_J$, which suggests 
that some singularity will emerge in the $T \to 0$ limit,
to be discussed below. Here, we simply emphasize that for the harmonic
sphere system, temperature must be at least lower than $T = 10^{-7}$
(in units of the spring constant $\epsilon$)
to notice the emergence of a singular volume fraction
dependence of the DW factors near the jamming transition. 
For larger temperatures, the DW factor smoothly depends on 
$T$ and $\phi$ as a result of the compression but the jamming 
singularity does not manifest itself. Anticipating 
the discussion of experimental work to come 
in Sec.~\ref{discussion}, we also remark that a singular behavior
can be seen in the data when the DW factors take extremely small
values, typically $\Delta^2 (\infty) < 10^{-4}$ (in units of the particle 
size), which corresponds to resolving 
distances with a precision greater than $\sigma / 100$, which can 
be technically quite challenging since it corresponds to 
about $10 {\rm nm}$ for colloids with $\sigma = 1 \mu{\rm m}$. 

\subsection{Velocity autocorrelation and density of states}

Another way of looking at single particle motion is to record the 
velocity autocorrelation function defined by
\be
d(t) = \frac{1}{3 N' T} \sum_{i=1}^{N'} \langle {\bf v}_i(t) \cdot 
{\bf v}_i(0) \rangle,
\label{dt}
\ee
where ${\bf v}_i(t)$ denotes the velocity of 
particle $i$ at time $t$.
Note that we again remove rattlers from the definition in 
Eq.~(\ref{dt}).
When Fourier transformed, this becomes what we (abusively) call 
the `density of states' (DOS)~\cite{oldtom}:
\be
d(\omega) = \int_0^\infty dt \cos(\omega t) d(t).
\ee
Note that $d(\omega)$ can be measured arbitrarily far from the 
harmonic limit and so it does not 
represent a genuine density of states, but 
it does reduce to the true DOS (as measured 
by diagonalizing the dynamical matrix) when
the harmonic approximation holds~\cite{dauchotmodes}. 
Below, we will carefully determine 
when this assumption is correct.
Note also that by definition one has 
\be 
d(t) = \frac{1}{2} \frac{\partial^2 \Delta^2(t)}{\partial t^2},
\ee
which shows that both types of measurements (density of states 
and mean-squared displacements) are directly related
and, in fact, contain nearly equivalent information. 
This remark suggests that, when carefully analyzed,
the MSD themselves should directly reveal the anomalous features of the 
DOS, as we shall demonstrate.

\begin{figure}
\psfig{file=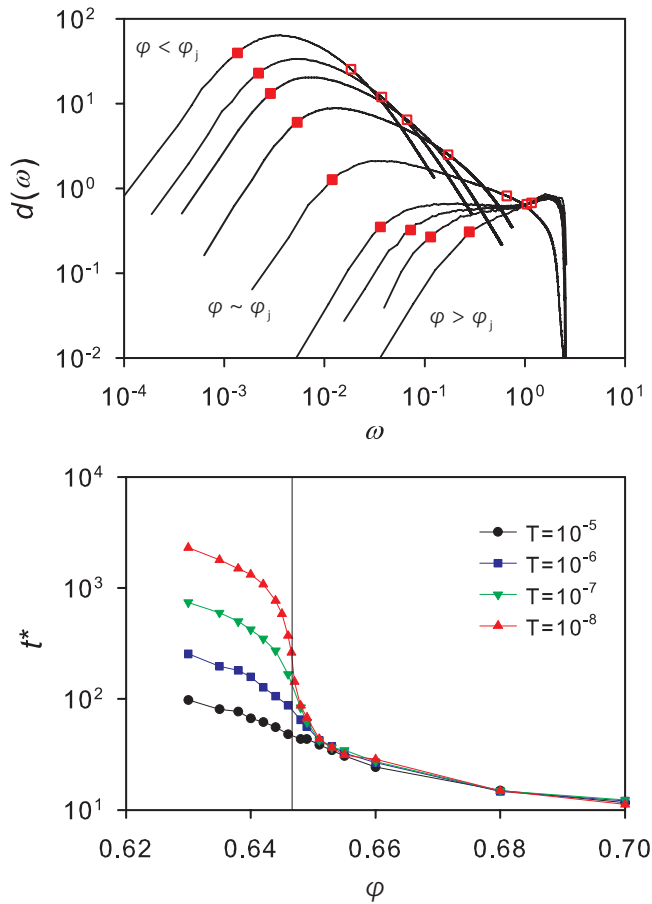,width=8.5cm,clip}
\caption{\label{fig2} 
Top: Frequency dependence of the density of 
states at constant temperature, $T=10^{-8}$, and densities
increasing from top to bottom.  Parameters are as 
in Fig.~\ref{fig1}.
Open squares indicate the microscopic frequency $\omega_0 = \pi /\tau_0$,
while filled squares indicate the low-frequency crossover 
$\omega^\star = \pi / t^\star$. 
Bottom: Volume fraction dependence of 
$t^\star = \pi /\omega^\star$
for different temperatures. The vertical line indicates $\phi_J$. } 
\end{figure}

In Fig.~\ref{fig2}, we present the frequency dependence of 
$d(\omega)$ at the same set of temperatures and 
densities as in Fig.~\ref{fig1}.
At low frequency, one gets the expected Debye behavior, 
$d(\omega) \sim \omega^2$ for our three-dimensional systems. 
Moving to larger frequencies, we see that
the DOS exhibits a single peak below jamming, $\phi < \phi_J$, and 
this peak broadens with increasing density and transforms into a plateau in 
the jammed regime, $\phi > \phi_J$. To determine the position of 
this low-frequency characteristic behavior over the entire 
density range, we define the crossover frequency $\omega^{\star}$
as the peak position of the quantity $d(\omega)/\omega$. In Fig.~\ref{fig2}
we indicate the frequency $\omega^\star$ with filled squares, 
which appear as good indicators of the low-frequency crossover 
towards the Debye power law. Clearly, $\omega^\star$ shifts to larger 
frequency when $\phi$ increases.  

At large frequency, $d(\omega)$ undergoes a second crossover towards
either an algebraic $d(\omega) \sim \omega^{-2}$ behavior below
jamming, or towards an even steeper decay above jamming. 
This behavior simply mirrors a qualitative change in the short-time 
behavior of the velocity autocorrelation function from exponential to 
Gaussian over the same density range~\cite{hansen}. 
We report the microscopic frequency 
$\omega_0 = \pi / \tau_0$ as open squares in Fig.~\ref{fig2}, which 
confirms that $\tau_0$ governs the high-frequency crossover 
in the DOS. 

The comparison between the MSD and DOS in Figs.~\ref{fig1} and 
\ref{fig2} confirms that both quantities contain 
equivalent information in the sense that they both reveal 
that vibrational dynamics is controlled by two distinct time scales,
$\tau_0 = \pi / \omega_0$ and $t^\star = \pi / \omega^\star$. 
The former indicates
when particles start to feel their neighbors and deviations 
from ballistic motion appear. The latter corresponds to the convergence
of the MSD to its long-time limit, and also to
a crossover to a Debye behavior in the DOS, meaning that
interesting dynamics arises in the range 
$\tau_0 \ll  t \ll t^\star$, or $\omega^\star \ll \omega \ll \omega_0$ 
in the frequency domain. 

Note also that the DOS reported in Fig.~\ref{fig2} 
are fully consistent with previous 
determination in both jammed~\cite{silbert} 
and unjammed regimes~\cite{brito}. The main feature noticed 
in previous work in these DOS data
is the emergence of the low-frequency crossover $\omega^\star$,
which we can alternatively interpret, in the time domain,
as the time it takes for the MSD to reach its plateau value,
see Fig.~\ref{fig1}.

We extract $t^\star = \pi / \omega^\star$ 
for a broad range of volume fractions and 
temperatures, see Fig.~\ref{fig2}.
As in the case of the DW factor, $t^\star$ shows very different temperature 
dependence in jammed and unjammed regimes. 
It is nearly independent of temperature in the jammed regime but scales
as $t^\star \sim \sqrt{T}$, reflecting the fact that temperature 
trivially renormalizes the unit time scale in hard sphere systems.
As temperature gets lower, we observe that $t^\star$ acquires a 
strong density dependence very close to $\phi_J$, changing by several orders 
of magnitude over a narrow density range. By contrast, 
$t^\star$ changes smoothly when $T$ is not low enough, 
as a consequence of the compression.

\section{Critical behavior}

\label{critical}

In the previous section, we discussed the qualitative behavior of 
various time scales characterizing the vibrational motion 
of dense packings of harmonic spheres. In particular, we noticed that
singular density dependences seem to emerge when temperature 
decreases. In this section, we focus on these 
low-temperature singularities and show that they correspond 
to zero-temperature algebraic divergences rounded by 
finite temperature effects, and 
therefore to an underlying critical dynamics.

\subsection{Elementary time scales and length scales}
\label{sec:tau0}

From the data reported in Figs.~\ref{fig1} and \ref{fig2}, one finds 
that the dynamics shows an abrupt crossover at low temperature 
when $\phi_J$ is crossed. Without further analysis, 
it is however not obvious to decide 
when a nontrivial critical behavior becomes manifest, as compression 
itself influences the dynamics. This suggests that one should  
first renormalize time scales and length scales in terms of the 
microscopic ones. In the case of the jamming transition, compared to 
usual critical phenomena, this is particularly 
important since the dynamics below and above jamming are very different:
dynamical behaviors are trivial for $\phi\ll\phi_J$ and  $\phi\gg\phi_J$ 
but in a quite distinct way. 

\begin{figure}
\psfig{file=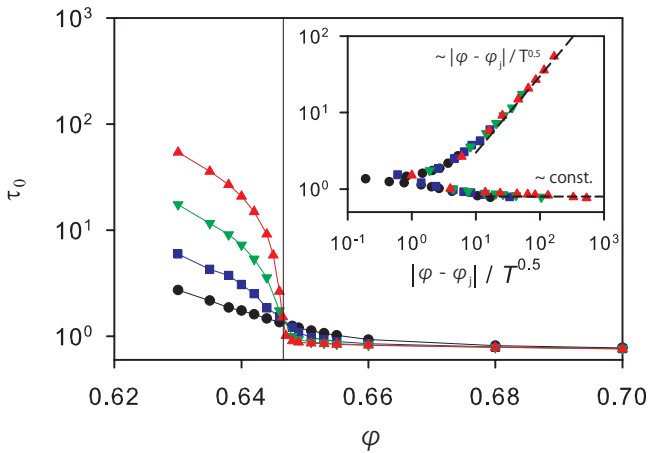,width=8.5cm,clip}
\caption{\label{fig3} Volume fraction dependence of 
the microscopic time scale $\tau_0$, Eq.~(\ref{deftau0}),
at various temperatures decreasing from bottom to top. 
The vertical line indicates $\phi_J$.
These data were used to draw the open squares in 
Figs.~\ref{fig1} and \ref{fig2}.
Inset: $\tau_0$ is replotted as a function 
of the scaling variable $|\phi-\phi_J|/\sqrt{T}$. 
Upper and lower dashed lines correspond to $T \to 0$ limits
for hard and soft spheres, respectively, 
Eqs.~(\ref{enskog}, \ref{einstein}).}
\end{figure}

Our first step is to analyze the short-time dynamics in order
to understand the behavior of the microscopic time scale
$\tau_0$ discussed above, which also sets the high-frequency cutoff
in the DOS $d(\omega)$.  In practice, we measure 
$\tau_0$ by analyzing the short-time decay of the 
velocity autocorrelation function $d(t)$ in Eq.~(\ref{dt}), 
and quantitatively determine $\tau_0$ from the definition 
\be 
d(t=\tau_0) = 1/e.
\label{deftau0}
\ee
The data for $\tau_0$ are shown in Fig.~\ref{fig3}.
These data were also used to determine the position of the open squares 
in the MSD data of Fig.~\ref{fig1}, and in the DOS data of 
Fig.~\ref{fig2}. 
 
The 
short-time behavior of the velocity autocorrelation function $d(t)$ 
is very different in jammed and unjammed regimes at low temperature. 
In the zero-temperature limit, $d(t)$ display an exponential decay when
$\phi < \phi_J$ due to two-body collisions in the hard sphere limit, 
whose characteristic time scale can be deduced from Enskog 
theory (roughly the time it takes to a particle to collide 
ballistically with a neighbor)~\cite{hansen}, 
\be
\frac{1}{\tau_0} \simeq \frac{8 \rho g(\sigma^+)}{3} 
\sqrt{ \frac{\pi T}{m}} 
\sim \frac{\sqrt{T}}{|\phi - \phi_J|}, 
\label{enskog}
\ee
where the final approximation only holds very close to jamming. 
We stress that this two-body collision time scale decreases 
very rapidly as $\phi_J$ is approached from below because the typical gap 
between neighboring particles also vanishes rapidly, 
scaling as $(\phi-\phi_J)$. This is quantitatively taken 
into account in Eq.~(\ref{enskog}) through the contact value 
of the pair correlation function, $g(\sigma^+)$.

In the low temperature limit above jamming, $\varphi > \varphi_j$, 
$d(t)$ shows instead a Gaussian decay, 
whose characteristic timescale can now be estimated 
as the inverse of the Einstein frequency~\cite{chaikin},
\be
\frac{1}{\tau_0} \simeq
\sqrt{ \frac{\rho}{3 m} \int d {\bf r} g(r) \nabla^2 v(r)} 
\sim \mbox{const}, 
\label{einstein}
\ee
where $g(r)$ is the pair correlation function, and 
the final approximation means that $\tau_0$ has no singular 
dependence upon either $T$ or $\phi$ near jamming.

The data in Fig.~\ref{fig3} indicate that $\tau_0$ interpolates 
smoothly between the two limits discussed above
on both sides of the transition at finite temperatures.
The main panel confirms that it is proportional to $\sqrt{T}$ below jamming, 
and becomes independent of temperature in the jammed regime. 
As in the case of $t^{\star}$, it is only in 
the very low temperature limit that a singular 
density dependence emerges. 
In the inset of Fig.~\ref{fig3}, $\tau_0$ 
is replotted as a function of the rescaled variable 
$x = |\phi-\phi_J|/\sqrt{T}$, in order to reconcile the behavior on
both sides, as is the case for most observables
near jamming~\cite{teitel,tom,hugo}.
Indeed we get a very good collapse of all the data along two 
distinct branches, confirming the simple dependences discussed above.
We emphasize that even though $\tau_0$ is controlled 
by relatively simple physics and it simply sets the 
microscopic time scale for the dynamics, it changes rather dramatically 
near jamming at low temperatures.

Having identified the elementary time scale $\tau_0$ on both sides
of the jamming transition, we shall now focus on the elementary length scale. 
Below jamming, as discussed above, particles move ballistically over the 
typical interparticle
distance in a time $\tau_0$. Thus, the elementary length scale obtained by 
ballistic motion reads:
\be
\ell_0=\sqrt{T}\tau_0.
\label{length}
\ee  
The same relation holds on the other side of the jamming transition. 
In that case the elementary length scale
is given by the typical displacement due to a harmonic mode with Einstein 
frequency (\ref{einstein}), leading again to 
Eq.~(\ref{length}). In both cases, $\ell_0$ represents
the distance travelled ballistically over the microscopic time
scale $\tau_0$.  

In conclusion, in this section we have identified the elementary 
length scales and 
time scales, Eqs.~(\ref{enskog}, \ref{einstein}, \ref{length}), 
which characterize noncollective dynamics.
In the next section we shall show 
that approaching the jamming transition, there emerge  
time scales and length scales which actually 
become much larger than $\tau_0$ and $\ell_0$, thus signalling
the existence of {\it bona fide} collective dynamics.

\subsection{Critical behavior of adimensional data}

We now reconsider the dynamical behavior studied in Sec.~\ref{dynamic}
and analyze the data obtained by renormalizing all observables 
in terms of the elementary time scales and length scales. 
 

We define a rescaled Debye-Waller factor as
\be
\Delta_\infty (\phi,T) \equiv  \frac{\Delta^2(\infty)}{\ell_0^2}
=\frac{\Delta^2(\infty)}{T \tau_0^2}.
\label{Delta}
\ee
and we similarly renormalize the long time scale $t^\star$  
in terms of $\tau_0$:
\be
\tau (\phi, T) \equiv  \frac{t^\star}{\tau_0}.
\label{tau}
\ee
A visual interpretation of these adimensional quantities
can be given from the symbols in Fig.~\ref{fig1}, since 
$\Delta_\infty$ and $\tau$ respectively quantify the vertical 
and horizontal distances between the filled and open squares,
i.e. between the short-time and long-time dynamics.
 
\begin{figure}
\psfig{file=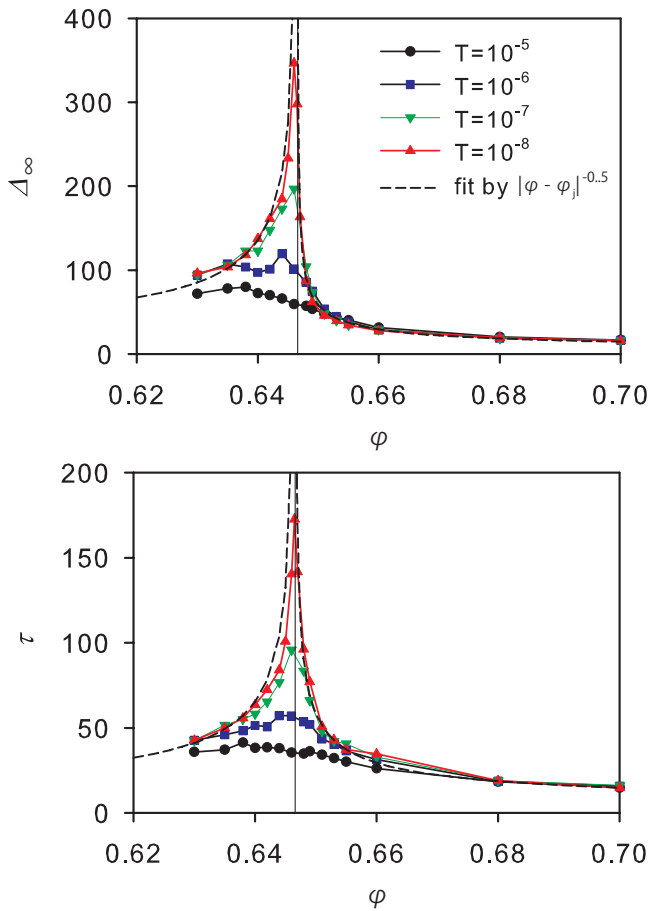,width=8.5cm,clip}
\caption{\label{fig4} Critical behavior for rescaled 
dynamic observables.
Top: Rescaled Debye-Waller factor, Eq.~(\ref{Delta}). 
Bottom: Rescaled time scale for vibrational motion, Eq.~(\ref{tau}). 
In both panels, dashed lines represent
the critical divergence $\sim |\phi - \phi_J|^{-1/2}$, Eq.~(\ref{powerlaw}), 
and the vertical line indicates $\phi_J$.}
\end{figure}

The numerical data for these two quantities are reported 
in Fig.~\ref{fig4}. 
Clearly, they both are smooth functions of the density 
when temperature is not very low (but still much smaller than 
the glass temperature), e.g. $T=10^{-5}$, and $T=10^{-6}$, and they 
also are smooth functions of the temperature when $\phi$ 
is not in the immediate vicinity of $\phi_J$. 
This implies that in this regime, the `low-frequency' crossover 
$\omega^\star$ is not very `low', i.e. not strongly decoupled 
from the natural microscopic time scale $\tau_0$. In the similar vein,
in the non-critical regime the DW factor is not very 
much larger than its natural scale,
and thus thermal vibrations are neither `anomalous' 
nor characterized by `soft' modes which would reflect 
an underlying nontrivial or collective dynamics. 

The situation becomes more interesting at very low temperature, 
$T \leq 10^{-7}$ and for densities near jamming, $0.64 \leq  \phi 
\leq 0.655$,
because adimensional quantities may become very large, 
and additionally they acquire a nonmonotonic density dependence 
across $\phi_J$, strongly  
suggesting the existence of divergences as $T \to 0$
on both sides of the transition. 
Thus, we conclude that the anomalous `low-frequency' behavior of the DOS
approaching jamming corresponds to the appearance of 
an adimensional timescale $\tau$ that becomes large, $\tau \gg 1$, and
an `anomalously' large adimensional amplitude of the vibrational motions, 
$\Delta_\infty \gg 1$. Thus, when analyzed from the viewpoint of thermal
vibrations, the jamming transition is {\it accompanied by a 
critical slowing} down which we now confirm through a scaling analysis.
In Sec.~\ref{lengthscale}, 
we will show that this slowing down is also 
accompanied by a diverging correlation length.

\subsection{Critical behavior and scaling in the $T \to 0$ limit}

We now connect the two critical behaviors found above and estimate the 
corresponding $T \to 0$ divergences
using a scaling analysis. This analysis amounts to assuming 
that  slow vibrational motion is controlled by a 
single dominant timescale, $t^*$, {\it i.e.} that the 
DOS at low frequency has a non-trivial scaling behavior 
of the form
\be
d(\omega)=\tau_0 f_{\pm} \left(\frac{\omega}{\omega^\star} \right), \qquad 
\omega \ll 1/\tau_0,
\label{scaling}\ee
where $f_{\pm}(x)$ corresponds to the universal behavior of the density of 
state below ($-$) 
and above ($+$) jamming, respectively.
We checked that numerical data are consistent with this scaling assumption.
Both functions $f_{\pm}(x)$ are 
proportional to $x^2$ for $x\rightarrow 0$ in order to 
recover the usual Debye law.  
For $x\rightarrow \infty$, $f_+(x)$ tends to a constant to recover the 
flat part of the density of states, while 
$f_-(x)$ decreases to reproduce the hard sphere behavior~\footnote{Note 
that these arguments hold even with less restrictive 
assumptions; one 
simply needs that $f(x)$ does not increase faster than $x$ when 
$x \to \infty$, but the specific form of $f_\pm(x)$ is irrelevant.}.

By definition of the MSD and of the DOS we get 
\be
\Delta^2 (t)  = 6  T 
\int_0^\infty d\omega \frac{d(\omega)}{\omega^2} [1- \cos (\omega t)].
\ee
In the long-time limit, the scaling behavior 
in Eq.~(\ref{scaling}) yields 
\be
\Delta^2 (\infty) \approx  
\frac{T\tau_0}{\omega^\star} \int_0^\infty dx \frac{f_{\pm}(x)}{x^2},
\label{fpm}
\ee 
which simply implies that 
\be 
\Delta_\infty \sim \tau.
\label{diff}
\ee
This result unveils that, to leading order, the data 
in both panels of Fig.~\ref{fig4} are in fact equivalent, 
up to a numerical prefactor stemming from 
the difference in the scaling functions $f_{\pm}(x)$ expressed 
by Eq.~(\ref{fpm}).
The result in Eq.~(\ref{diff}) has a simple physical interpretation,
because it says that a large amplitude for vibrational motions
develops near jamming, and that the time it takes for particles
to fully explore these vibrational degrees of freedom also 
becomes larger, the scaling between (adimensional) times and lengths 
obeying a simple diffusive scaling.
 
Finally, by using literature results concerning the scaling of $\omega^\star$ 
approaching the jamming transition at $T=0$~\cite{brito,wyart2}, we find that 
$\tau$ diverges symmetrically on both sides of $\phi_J$ as $|\phi - 
\phi_J|^{-1/2}$. 
For our rescaled variables, this implies that when $T \rightarrow 0$
\be 
\Delta_\infty \sim \tau \sim |\phi - \phi_J|^{-1/2},
\label{powerlaw}
\ee
on both sides of the jamming transition.
These scaling laws are in good agreement with our data, as shown 
with dashed lines in Fig.~\ref{fig4}. 
We have not tried to measure the exponent in Eq.~(\ref{powerlaw})
directly to detect a possible deviation from the 
value $-\frac{1}{2}$, but our numerical results 
suggest that this correction, if it exists, is quite small, 
in agreement with recent numerical work~\cite{brito,hayakawa}.

As is clear Fig.~\ref{fig4}, the divergence is cut off
by using a finite temperature. As we shall discuss 
later, this is due to the appearance of anharmonicity at finite temperatures.

\section{Diverging correlation length}

\label{lengthscale}

In the previous section, we found that 
the amplitude of the vibrational motion, and the associated
time scale, may become nontrivial, i.e.  much larger 
than the expected elementary values corresponding to non-collective 
dynamics. It is therefore natural to expect some kind of diverging correlation
length scale. It is the purpose of this section 
to unveil and study this correlation length. 

\subsection{Four-point correlation function}

Since we identified the mean-squared displacement as 
the quantity of interest to pinpoint the existence of the transition, 
we now seek spatial correlations between the particle displacements 
during the vibrational motion. 
This amounts to measuring the extent of spatial correlations 
in the dynamics of glassy states. We can therefore use the 
machinery developed to study dynamic heterogeneity in glassy 
materials~\cite{reviewchi4}.
First, we measure the following four-point susceptibility 
\be
\chi_4(t) = N \left[ \langle \mu(t)^2 \rangle 
- \langle \mu(t) \rangle^2 \right],
\label{defchi4}
\ee
defined as the variance of the spontaneous fluctuations 
of the instantaneous value of the 
volume-averaged mobility $\mu(t)$, defined as 
\be
\mu(t) = \frac{1}{N} \sum_{i=1}^N \frac{|\Delta {\bf r}_i(t)|^2}{\langle 
|\Delta {\bf r}_i(t)|^2 \rangle} - 1.  
\label{mu}
\ee
Note that the thermal average in the denominator 
of Eq.~(\ref{mu}) does not include 
an average over particles, and thus particles with larger displacements 
do not dominate the sum in the definition of $\mu(t)$. 
This implies in particular that rattlers are not dramatically affecting 
the numerical measurements of $\chi_4(t)$. 

While $\chi_4(t)$ measures the total amplitude of spatial correlations, 
direct access to a correlation length scale is provided 
by the spatially resolved correlator, which we study in the
Fourier space:
\be
S_4(q,t) = \frac{1}{N} \sum_{j,k = 1}^N 
\langle \left| \mu_j(t) \mu_k(t) e^{i {\bf q} \cdot 
{\bf r}_{jk}} \right| \rangle,
\label{S4}
\ee
where ${\bf r}_{jk} = {\bf r}_j(0) - {\bf r}_k(0)$ 
is the separation between particles $j$ and 
$k$ at $t=0$ and 
$\mu_j(t)$ is the contribution of particle $j$ 
to the mobility, 
$\mu_j(t) = |\Delta {\bf r}_j(t)|^2 / \langle |\Delta {\bf r}_jt)|^2 
\rangle - 1$.  

Two remarks 
before showing our results.
First, we have included all particles in the analysis of the dynamic 
heterogeneity, because the rattlers do not dominate 
the sum in Eqs.~(\ref{defchi4}, \ref{S4}), contrary to the MSD.
We have numerically checked that our results in this section 
are identical if we remove the rattlers.
Second, our definition of $\chi_4(t)$ does not explicitly 
involve a `probing' length scale as in other works~\cite{reviewchi4,leche}. 
In fact, using the mobility defined in Eq.~(\ref{mu}) 
is equivalent to selecting automatically the appropriate 
probing length at each time~\cite{leche}.

We find that $\chi_4(t)$ increases with $t$ at short times, but  
it does not display a clear maximum at intermediate times as is usually
found in systems near a glass transition~\cite{cristina},
but saturates to a long-time limit discussed below.
Here we wish to characterize the spatially heterogeneous dynamics 
relevant for $\Delta_\infty$ and $\tau$. 
We therefore concentrate on time scales of the order of $t^\star$ discussed
above, and we measure 
\be 
\chi_4(\phi,T) \equiv \chi_4(t=t^\star), 
\label{chi4}
\ee
and the corresponding structure factor: 
\be 
S_4(q;\phi,T) \equiv S_4(q,t=t^\star).
\label{s4}
\ee

\begin{figure}
\psfig{file=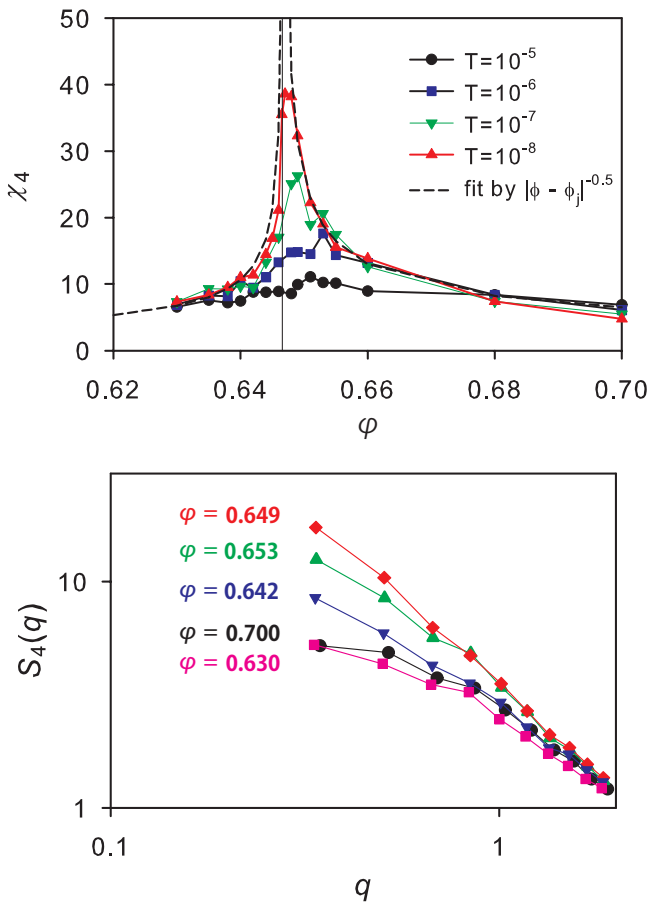,width=8.5cm,clip}
\caption{\label{fig5} 
Top: Volume fraction dependence of the dynamic 
susceptibility $\chi_4(t=t^\star)$, Eq.~(\ref{chi4}), for various 
temperatures showing the emergence of `vibrational heterogeneity'
at low enough $T$ close to $\phi_J$ (shown with the vertical line). 
Bottom: Four-point correlator $S_4(q,t=t^\star)$ at $T=10^{-8}$ 
and various volume fractions exhibits a nonmonotonic density 
dependence (note the nonmonotonic evolution of the $\phi$ labels).}
\end{figure}

Representative numerical results are shown in Fig.~\ref{fig5}. 
As for the adimensional quantities $\Delta_{\infty}$ and $\tau$
in Fig.~\ref{fig4}, $\chi_4$ also shows a striking
nonmonotonic density dependence across $\phi_J$, 
which becomes more pronounced as $T$ decreases, and would seem 
to diverge on both sides of the jamming transition as $T \to 0$.
These results strongly suggest that the slowing down and excess 
amplitude of thermal vibrations are the result of collective, 
spatially correlated dynamics. Remark again that 
$\chi_4$ acquires a strong nonmonotonic density 
dependence in the close vicinity of $\phi_J$, but only for temperatures 
which are at least below $T \approx 10^{-7}$, 
while it remains nearly constant and rather insensitive to the 
underlying jamming transition everywhere else.  
Finally, remark that the critical behaviour of $\chi_4$ emerges in 
Fig.~\ref{fig5} without further analysis or normalization 
by an elementary `correlation unit'~\footnote{It would not make sense to 
renormalize the wavevector 
in $S_4(q)$, and thus $\chi_4$, in terms of $\ell_0$ since the dynamic 
correlation length corresponds to the number of spatially 
correlated particles and is not related to the scale 
over which single particles move.}.  
This makes it a very relevant quantity to be measured experimentally, 
as it directly reveals the dynamic criticality associated to 
the jamming transition.

We now focus on spatial correlations to extract 
the corresponding correlation lengh scale over which 
thermal vibrations occurring over a time scale $t^\star$ are correlated.   
In Fig.~\ref{fig5}, we present the evolution of 
$S_4(q)$ for $T=10^{-8}$ across the jamming density. 
As suggested by the behavior of $\chi_4$, we find that 
$S_4(q)$ is also characterized by a nonmonotonic behavior 
with $\phi$, spatial correlations being modest at both 
low and large $\phi$, and being maximum near $\phi_J$.
Note that $S_4(q \to 0)$ represents the volume integral 
of the spatial correlation between particle displacement, and it 
is thus natural that the low-wavevector limit of $S_4(q)$ follows
the same trend as $\chi_4$~\cite{jcp}. 
However, the spatially resolved correlator
allows us to directly extract a correlation length through the 
$q$ dependence of $S_4(q)$. Clearly, the approach to the low-$q$ plateau
occurs for lower $q$ when $\phi \approx \phi_J$, which indicates that
the corresponding correlation length is also maximum there
at finite $T$, and would diverge at $\phi_J$ in the
$T \to 0$ limit.

Before discussing the divergence at $T=0$ in more detail, 
we wish to emphasize the striking similarity between the
numerical results displayed in Fig.~\ref{fig5} quantifying the 
the `vibrational heterogeneity' of our thermally excited packings 
of soft particles, and a similar nonmonotonic  dynamic heterogeneity
reported for gently vibrated grains in Refs.~\cite{leche,corentin}. 
Although the excitation mechanisms and the physical 
systems are quite distinct, it is very tempting 
to suggest that both sets of measurements are actually 
probing the same underlying dynamic criticality associated to the 
jamming transition. 

\subsection{Critical behavior and scaling in the zero temperature limit}

We now connect these observations of spatially heterogeneous
vibrational dynamics  
with the critical behavior of $\tau$ and $\Delta_{\infty}$, 
extending the previous scaling analysis.
Our starting point to evaluate $\chi_4$ is to assume that 
fluctuations are Gaussian and, hence, four-point correlation functions 
can be written as products of two-point functions. This is correct 
within the harmonic 
approximation, which is valid in the zero-temperature limit we are 
considering for this analysis. 
With this assumption, we can obtain an analytical 
expression of the long-time limit, $t \to \infty$, of the 
four-point susceptibility as 
\be 
\chi_4(t\to\infty) \approx \Bigl(\frac{T}{\Delta^2(\infty)}\Bigr)^2 
\int_{\omega_{\rm min}} d \omega 
\frac{d(\omega)}{\omega^4}. 
\ee
Here $\omega_{\rm min}$ is the lowest frequency of the density of state 
for a given system. 
When the system size becomes large, $\omega_{\rm min} \sim L^{-1} \to 0$ 
and $d(\omega) 
\propto \omega^2$ for small $\omega$, 
and then, the integral displays an infrared divergence
which is cutoff at a system size dependent value, 
as we have numerically confirmed. 
This is because the lowest frequency mode is a plane wave for which 
the motion of all the particles in the system are fully 
correlated~\cite{cristina}.
 
We focus instead on finite times, $\chi_4 = \chi_4 (t=t^*)$:
\be 
\chi_4 (t^*) \approx \Bigl(\frac{T}{\Delta^2(t^*)}\Bigr)^2 \int_{\omega^\star} 
d \omega 
\frac{d(\omega)}{\omega^4}, 
\ee
from which we deduce
\be 
\chi_4 \sim \frac{1}{d(\omega^\star) \omega^\star} \sim \tau. 
\ee
Finally, if we assume (to be confirmed in a short while) an Ornstein-Zernike
form for the correlation length we also obtain 
the relation between susceptibility and length scale, 
$\xi_4 \sim \sqrt{\chi_4}$.

To summarize our scaling analysis, we have shown that the following 
quantities are all connected in a simple way at the level of 
their scaling behaviors, and we obtained the 
following critical scalings:
\be 
\Delta_\infty \sim \tau \sim \chi_4 \sim \xi_4^2 .
\label{exp2}
\ee
Note that in order to derive these results we only used the harmonic 
approximation (valid when $T \to 0$) and 
the scaling assumption (\ref{scaling}). If we add the information, 
obtained numerically in Refs.~\cite{brito,wyart2}, 
that $\tau$ diverges as $|\phi-\phi_J|^{-1/2}$ then we also get the zero 
temperature divergence of 
$\Delta_\infty\propto |\phi-\phi_J|^{-1/2}$ and of $\xi_4\propto |\phi-
\phi_J|^{-1/4}$ approaching the jamming transition.
As discussed before and is clear from our data, any finite temperature 
cuts off these divergences.

\begin{figure}
\psfig{file=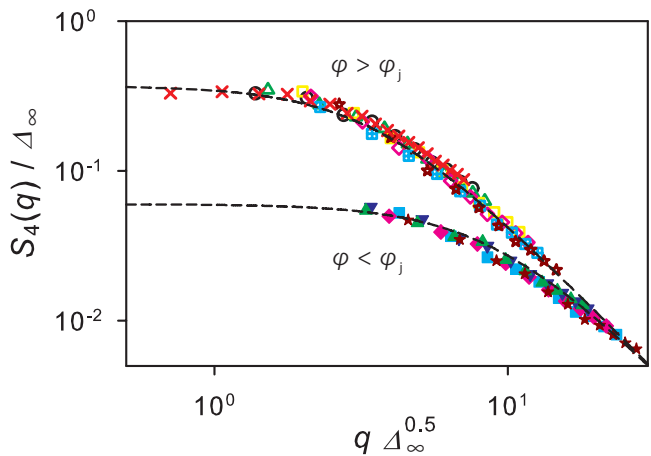,width=8.5cm,clip}
\caption{\label{fig6} 
Test of the scaling behavior of $S_4(q)$ using the 
assumption that $\xi_4 \sim \sqrt{\Delta_\infty}$.
Note that this data collapse involves no free parameter
since $S_4$ and $\Delta_\infty$ result from indepent measurements.
Dashed lines are fits to the Ornstein-Zernike functional form,
Eq.~(\ref{OZ}).}
\end{figure}

According to our analysis, the diverging correlation 
length is proportional to the square root 
of the diverging excess amplitude of the thermal vibrations, 
$\xi_4 \sim \sqrt{\Delta_\infty}$.  
We directly test this hypothesis in Fig.~\ref{fig6} where we
present $S_4(q)/\Delta_\infty$ as a function of $q\sqrt{\Delta_\infty}$. 
Clearly, all the data are collapsed very well along two branches
corresponding to both sides of the jamming transition. 
We emphasize that this data collapse involves no adjustable 
parameter, but makes use of two independent sets of measurements.
Additionally, we show that
both branches are correctly described by an Ornstein-Zernike
wavevector dependence, namely
\be 
S_4(q) \approx \frac{ c_{\pm} \Delta_\infty}{1+ c_{\pm} q^2 \Delta_\infty},
\label{OZ}
\ee
with $c_{\pm}$ a numerical prefactor which is distinct for the 
two branches. 
These data directly confirm the validity of 
the analysis in Eq.~(\ref{exp2}).

We conclude that the appearance of `anomalously' slow and large 
vibration motion is also associated with spatially heterogeneous dynamics,
which is a form of {\it dynamic criticality}~\cite{rmp77}.
These observations indicate that the jamming transition
appears as a genuine critical transition with both a critical 
slowing down and a diverging length scale. Since the 
slowing down occurs for thermal vibrations, this suggests that
a convenient order parameter to describe the transition 
is the vibrational motion within cages deep into the glassy phase.
It is intriguing that all the critical exponents describing 
time scales, amplitude of vibration, 
and dynamic heterogeneity are related in a trivial manner,
Eq.~(\ref{exp2}). Interestingly these divergences are symmetric and 
hold quantitatively for both hard and soft spheres, provided
adimensional quantities are considered. 

We note that the length scale quantifying the spatial correlations
in the excess amplitude of the vibrations diverges, for 
the harmonic potential used in this study, as 
\be
\xi_4 \sim |\phi - \phi_J|^{-1/4}.
\label{length2}
\ee
A length scale diverging with the same exponent was already discussed
in different contexts. In Ref.~\cite{silbert}, a diverging length 
is identified from the transverse structure factor of the normal modes
obtained for soft spheres above jamming. This length scale 
also appears in the analysis of Ref.~\cite{vitelli2} of the heat 
transport in jammed solids. Finally, the same length scale 
also shows up in a mean-field treatment of the localization
length of the elastic response of model networks~\cite{recentwyart}. 
To our knowledge, for hard 
spheres, this length scale was not discussed before. Surprisingly, in both 
soft and hard
cases, the isostatic length scale introduced in Ref.~\cite{wyart2} 
does not directly appear in any of the physical quantities 
we discuss in this work. 
Finally, $\xi_4$ in Eq.~(\ref{length2})
seems to differ also from the correlation lengths measured 
in Refs.~\cite{teitel,hatano,drocco}. 

Remark that in our scaling analysis, arguments  
involving the average contact number are not directly employed. 
It is only to obtain the divergences of 
$\tau$, $\Delta_\infty$, and $\xi_4$ with $|\phi-\phi_J|$ that we 
made use of the scaling of $\omega^\star$ with $|\phi-\phi_J|$
which was related to the distance to isostaticity~\cite{wyart2}. 
We stress that none of those divergences is fully 
understood theoretically from a microscopic viewpoint. In fact, 
isostatic arguments~\cite{wyart2} trace back the behavior of $\omega^*$ 
with $|\phi-\phi_J|$ 
to the one of the contact number, 
which is not directly 
derived, but measured from numerical simulations. 
An alternative theoretical approach, which is fully microscopic, uses
statistical mechanics and replica calculations 
to study the jamming transition~\cite{PZ}.
When applied to harmonic spheres~\cite{hugopre,hugo}, it
correctly predicts that the jamming transition 
is characterized by a diverging amplitude of thermal vibrations: 
$\Delta_\infty \sim |\phi - \phi_J|^{-1}$.
However, this predicted divergence is only qualitatively correct, and 
the predicted critical exponent is wrong, see Eq.~(\ref{exp2}). 
It would be interesting to understand better the source 
of this discrepancy from the replica viewpoint~\cite{jorge}. Our results
also suggest that the theory should be reconsidered in order to 
make precise predictions for a diverging correlation length and 
dynamic susceptibility associated to the jamming transition.

\section{Finite temperature critical behavior}

\label{harmonic}

In the following we will fully describe the critical behavior in the 
$(\phi,T)$ phase diagram at the jamming transition. 
We will first analyze the low-$T$ harmonic regimes in both 
jammed and unjammed phases. We shall show that their domain 
of validity is 
restricted to a temperature regime that shrinks approaching $\phi_J$. 
This corresponds 
to an emergent temperature scale, $T^\star(\phi)$, that vanishes 
quadratically with $|\varphi - \varphi_J|$. 
Finally, we shall focus on the finite temperature critical regime, 
corresponding to $T^\star\ll T\ll1$, 
that is related to the critical divergences when 
$T\rightarrow 0$ for $\phi=\phi_J$. 

\subsection{Regime I: Harmonic description above jamming and its 
domain of validity}

In the following we shall establish that approaching the jamming transition 
from the jammed phase, the harmonic description used in the previous section 
is valid and allows one to explain the behavior of physical observables 
related to vibrational motions if temperature is low enough.
Moreover, we shall find the temperature scale, $T^\star (\phi)$, above which 
the harmonic description breaks down and the low temperature physical
behavior changes. 

\begin{figure}
\psfig{file=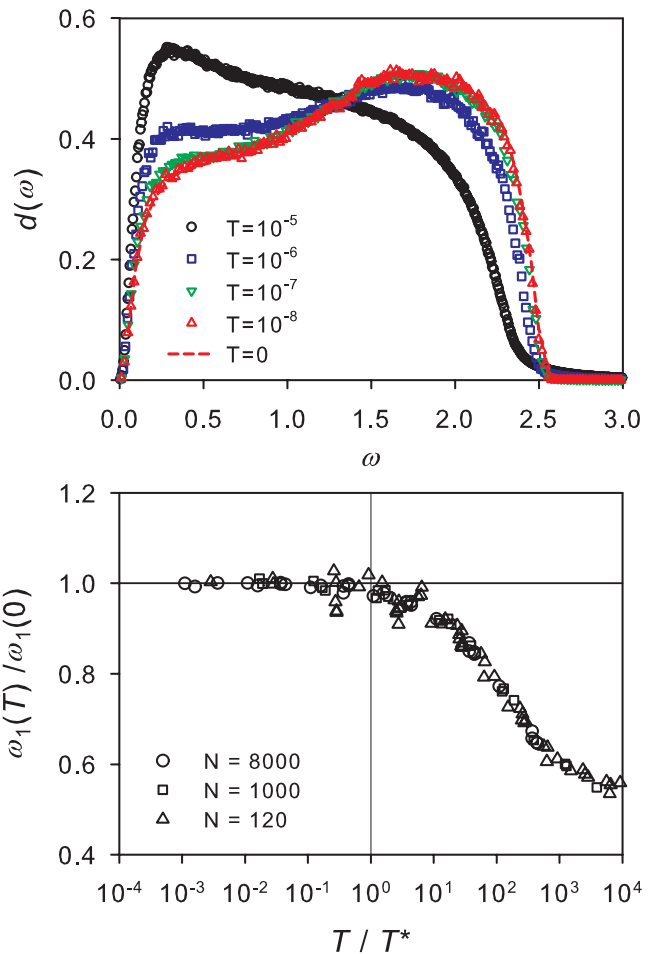,width=8.5cm,clip}
\caption{\label{fig7} 
Limit of validity of harmonic approximation above jamming.
Top: The density of state $d(\omega)$ at $\varphi=0.655$ 
($\phi - \phi_J \approx 0.0084$) measured 
for various temperatures through Eq.~(\ref{dt}).
The $T=0$ DOS is obtained from diagonalization of the dynamical matrix.
Bottom: Convergence of the first moment of the density of state
to its $T = 0$ limit below a crossover temperature scale 
$T^\star = 10^{-3} (\phi - \phi_J)^2$ which is independent
of system size. The vertical line indicates $T/T^\star = 1$.}
\end{figure}

To this end, we first focus on the temperature dependence of the DOS measured
via Eq.~(\ref{dt}) for $\phi > \phi_J$.  
Typical data are shown in Fig.~\ref{fig7} together with $T=0$ result 
obtained from the diagonalization of dynamical matrix in 
the ground state at $T=0$. 
We observe that by increasing the temperature, deviations from the 
$T=0$ result emerge and become more pronounced. Notice, in particular,
that it is near the high-frequency cutoff that deviations appear 
first when $T$ increases. By normalization, the low-frequency 
part of the DOS is then also affected.
Qualitatively similar results were shown in Ref.~\cite{xu}.

For the particular density shown in Fig.~\ref{fig7}, we observe  
that the DOS converges to its $T \to 0$ limit when 
$T < T^\star \approx 10^{-7}$. Thus, by contrast to the results 
of Ref.~\cite{ohern} we conclude that {\it a finite 
temperature region exists} where the harmonic approximation holds. 

To quantify the size of this harmonic regime, we 
measure the changes in $d(\omega)$ through its first moment: 
\be
\omega_1(\phi,T)  = \int_0^\infty d\omega \, d(\omega) \omega.
\ee
The observed change in $d(\omega)$ suggest that $\omega_1$ should 
decrease with increasing $T$, because the DOS gives less weight to 
large frequencies when $T$ increases.

Our numerical analysis fully confirms this expectation, see Fig.~\ref{fig7}.
We find that, for any given  density $\phi > \phi_J$, 
the first moment $\omega_1$ decreases above some crossover 
temperature $T^\star$, which depends strongly on the volume fraction. 
To see this, we rescale in the lower panel of 
Fig.~\ref{fig7} the first moment $\omega_1$ by its value at 
$T=0$, and present its evolution as a function
of the rescaled temperature $T/T^\star$. 
Our numerical results are consistent with a simple behavior 
\be
\frac{\omega_1(\phi,T)}{\omega_1(\phi,T=0)} \approx \mathcal{W} 
\left( \frac{T}{T^\star} \right),
\ee 
where $\mathcal{W}(x \to 0) = 1$, with 
\be
T^\star \approx a ( \phi - \phi_J)^2, \label{temp1}
\ee 
where $a \simeq 10^{-3}$ is a numerical prefactor.
There is an ambiguity in the absolute value of the 
numerical coefficient $a$, 
since the scaling collapse in Fig.~\ref{fig7} does not depend on it. 
We have determined $a$ such that 
$\mathcal{W}(x)$ starts to show deviations from 1 when $x \approx 1$.

To fully demonstrate that a harmonic regime exists in the thermodynamic
limit, we must make sure that $T^\star$ 
does not decrease to zero as the system size
$N$ is increased. Indeed, a strong system size dependence of 
the onset of anharmonicity was reported in Ref.~\cite{ohern}.
To test this idea, we have repeated the above measurements for various 
system sizes, $N=120$, 1000, and 8000. As shown in Fig.~\ref{fig7} 
we find that our results are devoid of any system size dependence,
although the statistical noise is of course larger for smaller systems. 

The fact that $T^\star$ vanishes quadratically with $(\varphi - \varphi_J)$
for all system sizes confirms that materials exactly at $\phi_J$ are 
`marginally' solid with special properties.
In particular, the linear response and harmonic approximation 
regimes have a vanishing domain of validity~\cite{martin,vitelli}.
Our results seem consistent with the analysis of low-energy barriers 
performed numerically in Ref.~\cite{vitelli}.
In particular, we conclude that 
the $T=0$ DOS cannot be used at $\phi_J$ to infer 
properties of the vibrational dynamics at any finite temperature. 
This resolves the problem raised in the introduction. 
The material does not `melt' at jamming when thermal fluctuations 
are added, because a finite amount of thermal noise
pushes the system in a anharmonic regime preventing
the system from breaking apart. 

\begin{figure}
\psfig{file=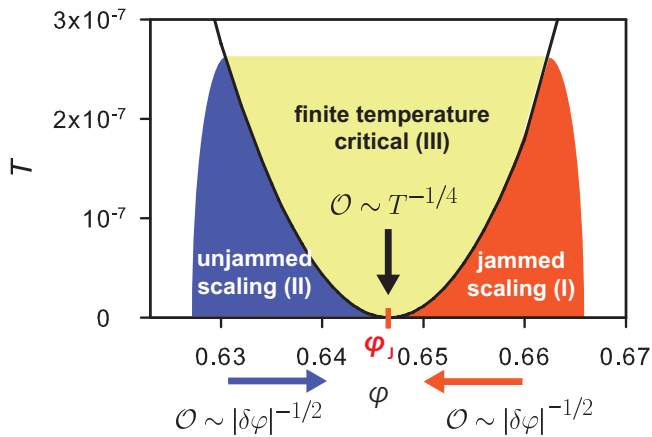,width=8.5cm,clip}
\caption{\label{fig10a} 
The structure of the three scaling regimes where
power law divergences and `anomalous' vibrational motion is
observed. Regimes I and II are described by $T=0$ harmonic 
theories, while dynamics in Regime III is fully anharmonic.
These regimes are separated by the crossover 
temperature $T^\star \sim 10^{-3} (\phi - \phi_J)^2$
discussed in Sec.~\ref{anharmonic}. Note the small range of 
parameters where the effect of the $T=0$ 
jamming critical point are felt and Eqs.~(\ref{div2}, \ref{div3}) hold.} 
\end{figure}

In conclusion, we find that for  $\phi>\phi_J$ and $T<T^\star(\phi)$ the 
density of states converges to its 
zero temperature limit which is correctly described by the harmonic 
approximation. In this regime the critical
behavior is the one discussed previously in the zero 
temperature limit.
We shall call this part of the phase diagram `Regime I', 
see Fig~\ref{fig10a}.
Above the temperature scale  $T^\star(\phi)$ the harmonic approximation 
breaks down and 
the critical behavior crosses over to Regime III that we discuss 
later. An explanation for the quadratic scaling of $T^\star(\phi)$ is
presented in Sec.~\ref{anharmonic}.

\subsection{Regime II: Effective harmonic description 
below jamming and its domain of validity}

Below jamming, the $T \to 0$ limit inherently produces 
a strongly anharmonic system, because it produces 
hard sphere packings with zero energy, and the Hessian is not 
defined. However, an effective harmonic description that 
describes this behavior for $\phi<\phi_J$ 
can be approximately obtained~\cite{brito}. 
In this section we repeat the analysis 
performed in Regime I to determine the limit 
of validity of the effective harmonic description 
below jamming, which defines `Regime II' in the phase diagram,
see Fig.~\ref{fig10a}. 

\begin{figure}
\psfig{file=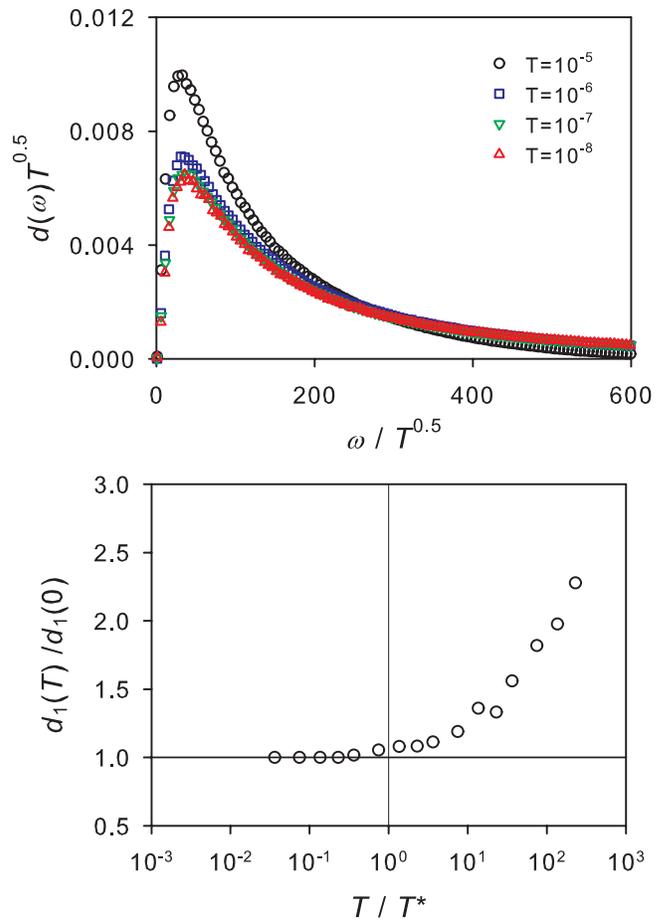,width=8.5cm,clip}
\caption{\label{fig8} 
Limit of validity of (effective) harmonic approximation below jamming.
Top: Temperature-scaled density of state $d(\omega)\sqrt{T}$ at 
$\varphi=0.630$ and various temperatures converges
to $T = 0$ limit below $T^\star \approx 10^{-7}$.
Bottom: Convergence of the height of the first peak of the DOS 
to its $T=0$ value below the crossover temperature scale 
$T^\star(\phi) = 10^{-3} (\phi_J - \phi)^2$. 
The vertical line indicates $T/T^\star = 1$.}
\end{figure}

In Fig.~\ref{fig8}, 
we characterize the temperature evolution of the DOS at fixed volume 
fraction and decreasing temperature at $\phi=0.63 < \phi_J$. 
To remove a trivial temperature dependence of the time scale 
for hard spheres, we plotted $d(\omega)/\sqrt{T}$ versus $\omega \sqrt{T}$,
as it is this rescaled DOS which eventually becomes 
independent of $T$ as $T \to 0$. 
 
Again, we observe that the DOS actually converges to its $T \to 0$ 
limit for this particular density below $T^\star \approx 10^{-7}$. 
To quantify the extent of the hard sphere regime, we focus 
on the height of the first peak of the DOS, $d_1(T)$,
which is a very sensitive indicator of the 
temperature evolution of $d(\omega)$. 
As in the case of the first moment above jamming, we follow 
the convergence $d_1(T)$ to its $T = 0$ value, and determine 
the crossover temperature $T^\star(\phi)$ controlling this convergence,
see Fig.~\ref{fig8}.
The numerical results show that there is a again a simple behavior
for this quantity,   
\be
\frac{d_1(\phi,T)}{d_1(\phi,T=0)} \approx 
\mathcal{D} \left( \frac{T}{T^\star} \right),
\ee 
where $\mathcal{D}(x \to 0) = 1$ with 
$T^\star$ taken from the previous section.
As in the case of jammed phase, there is a slight ambiguity 
in the actual prefactor of $T^\star$. For the unjammed region, 
we simply use the $a$ value determined for the jammed region, 
and find that this works very well. 

In conclusion, we find that for  $\phi<\phi_J$ and $T<T^\star(\phi)$ the 
density of states converges to its 
zero temperature limit which is correctly described by the effective 
harmonic approximation \cite{brito}. In this Regime II, the critical
behavior is the one discussed previously in the zero temperature limit,
see Fig~\ref{fig10}.
For $T> T^\star(\phi)$ the harmonic approximation 
breaks down and the critical behavior crosses over 
to Regime III that will be discussed in the following. 
An explanation for the quadratic scaling of $T^\star(\phi)$ will be 
presented in Sec.~\ref{anharmonic}.

\subsection{Regime III: Anharmonic critical regime 
at finite temperature}

We found in previous sections that below the temperature scale
$T^\star(\phi)$, which  
vanishes quadratically and symmetrically 
in $|\phi-\phi_J|$ approaching the jamming 
transition from both sides, 
the critical behavior is well described by the harmonic theory,
which leads to symmetric divergences on both
sides of $\phi_J$, captured by 
\be 
\Delta_\infty \sim \tau \sim \chi_4 \sim \xi_4^2 \sim |\phi-
\phi_J|^{-1/2}
\label{div2}
\ee
We also showed that a finite temperature 
cuts off these divergences. This means that 
at $\phi=\phi_J$ all these observables are finite
but diverge as $T \rightarrow 0$. This represents 
a distinct critical regime which cannot 
be described by the harmonic approximation, 
and which is therefore 
controlled by anharmonicity and finite temperatures. We call 
this `Regime III', see Fig~\ref{fig10a}.

In order to obtain scaling laws in this regime, we use scaling 
arguments and assume that the behavior of the physical
observables on both sides of the jamming transition is connected via 
(observable-dependent) scaling functions~\cite{tom,teitel}: 
\be
{\mathcal O}(\phi, T)=\frac{1}{|\phi-\phi_J|^{1/2}} 
{\cal H}_{\mathcal O}\left(\frac{\phi-
\phi_J}{\sqrt{T}} \right),
\label{scalingfunction}
\ee
where ${\mathcal O}$ stands for any of the following 
observables: $\Delta_\infty$, $\tau$,  $\chi_4$ or $\xi_4^2$. 
The three regimes I, II, and III described in the phase diagram in 
Fig~\ref{fig10} correspond respectively to $x \gg 1$, $x \ll -1$, 
and $|x|\ll 1$. Therefore, 
the scaling functions must be such that:
\be
{\cal H}_{\mathcal O}(x) 
\rightarrow  c_{\mathcal O}^{\pm}, \quad x \rightarrow \pm \infty,
\ee
in order to recover the correct $T=0$ divergences characteristic of 
regimes I and II, respectively; $c_{\mathcal O}^{\pm}$
are observable-dependent numerical constants.

The critical behavior in regime III can then 
be obtained by requiring that 
when $\phi \rightarrow \phi_J$ at small but finite temperature 
the dependence on $\phi-\phi_J$ drops out 
from Eq.~(\ref{scalingfunction}). 
In this way, one finds that at $\phi=\phi_J$:
\be 
\Delta_\infty \sim \tau \sim \chi_4 \sim \xi_4^2 \sim T^{-1/4}
\label{div3}
\ee

The critical behaviors of the three regimes 
are summarized in Fig.~\ref{fig10a}.
Note that contrary to regimes I and II, physical or microscopic 
interpretation for the divergences in Eq.~(\ref{div3}) are 
missing. A possible path could be the 
extension of the effective harmonic description 
of Ref.~\cite{brito} to continous potentials, but this is beyond the scope
of this work. 

\section{Anharmonicity and the temperature scale $T^\star$}

\label{anharmonic}

We have found by numerical simulations that the harmonic description 
for the jammed and unjammed phase
breaks down below a characteristic temperature scale vanishing 
quadratically with the distance to $\phi_J$. In the following we 
explore the origin of the anharmonicity responsible for this behavior 
and offer an explanation for the scaling of $T^\star$ with $|\phi-\phi_J|$.

\subsection{Perturbative analysis of high-order nonlinearities}

A first natural idea to explain the breakdown of the harmonic approximation 
is to realize that the harmonic analysis corresponds to a 
truncation to second order 
of an expansion of the energy function around a global minimum,
which is generally justified if the temperature is low enough. 
Thus, the emergence of anharmonicity at finite temperature 
could be due to higher order terms in the expansion 
becoming relevant. 

To explore this hypothesis, we focus on the jammed phase 
where a genuine harmonic approximation holds, and 
an expansion of the energy can be performed. To this end, 
we compute the first low-temperature corrections
to the average energy due to higher-order terms,  
and determine the temperature above which these
nonlinearities can no longer be neglected.

We expand the position vector of a particle 
using a normal mode decomposition defined from a given energy minimum.
We write 
\be 
{\bf r}_i = {\bf r}_i^{(0)} + \sum_a x_a {\bf n}_{a,i}, \label{rex}
\ee 
where ${\bf r}_i^{(0)}$ is the position of particle in the ground state and 
${\bf n}_{a,i}$ is the 
displacement of the $i$-th particle in the $a$-th normal mode. 
Using these normal modes, we can also expand the potential 
energy around the ground state energy into a power series of $\{ x_a \}$.  
Taking the thermal average of this series, 
we obtain the following temperature expansion form 
of the potential energy (details of this calculation are reported 
in the Appendix):
\be
\langle E \rangle = E_{\rm gs} + E_{1} T + E_{2} T^2 + 
\mathcal{O}(T^3),  \label{ptg}
\ee
where $E_{\rm gs}$ is the ground state energy, and 
\begin{eqnarray}
E_{1} &=& \frac{1}{2} \sum_{a,b} \frac{\partial^2 V}{\partial x_a 
\partial x_b} \frac{\delta_{ab}}{\lambda_a} = \frac{3N}{2} \label{pt1} \\
E_{2} &=& -  \frac{1}{8} \sum_{ab} \frac{1}{\lambda_a  \lambda_b} 
\frac{\partial^4 V}{\partial x_a^2 \partial x_b^2} \nonumber \\
&& + \frac{1}{8} \sum_{abc} \frac{1}{\lambda_a \lambda_b \lambda_c} 
\frac{\partial^3 V}{\partial x_a \partial x_b^2} \frac{\partial^3 
V}{\partial x_a \partial x_c^2} \nonumber \\
&& + \frac{1}{12} \sum_{abc} \frac{1}{\lambda_a \lambda_b \lambda_c}  \Bigl( 
\frac{\partial^3 V}{\partial x_a \partial x_b \partial x_c}\Bigr)^2. 
\label{pt2} 
\end{eqnarray}
From these results, we can estimate a characteristic 
temperature, $T_{\rm pt}$, at which the quadratic term 
in the temperature expansion becomes equal to the linear one, 
signalling that a harmonic expansion would break down:
\be
T_{\rm pt} = \frac{E_1}{E_2}.
\label{tpt}
\ee

\begin{figure}
\psfig{file=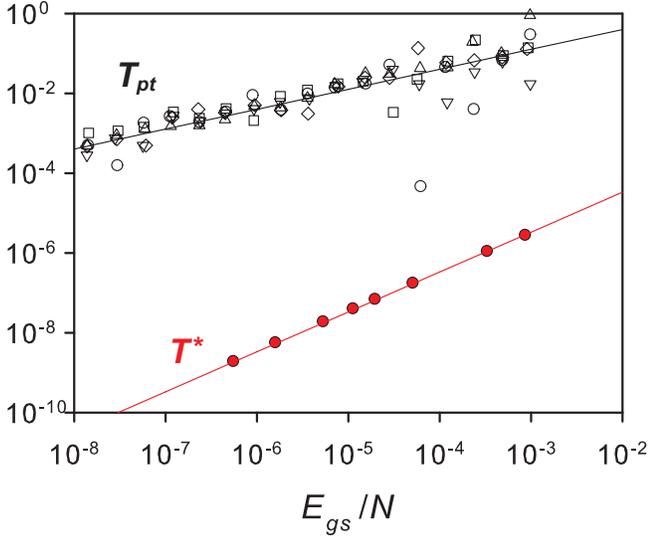,width=8.5cm,clip}
\caption{\label{fig11} 
The crossover temperature scale for the onset of anharmonicity
$T^{\star}$ (bottom data with filled circles) compared to the temperature scale 
$T_{\rm pt}$ (top data with open symbols different symbols for different 
packings) 
quantifying the relevance of the high order terms 
in the perturvative expansion of the energy.
The lines indicate $T^\star \propto E_{\rm gs}$ and $T_{\rm pt} \propto 
E^{1/2}_{\rm gs}$. Clearly, these two temperature scales
differ.}
\end{figure}

We have computed $T_{\rm pt}$ for 10 independent configurations 
with $N=256$ particles. Due to the multiple sums in 
Eq.~(\ref{pt2}) we are forced to use relatively small system sizes, 
but we have numerically checked that system size 
dependence is negligible, or at least much 
too weak to influence our conclusions.
For each configuration, we determine numerically the 
normal modes at $T=0$ for various densities above jamming.
From the numerical determination of the modes, we estimate 
$E_{1}$ and $E_{2}$ in Eqs.~(\ref{pt1}, \ref{pt2}), and we 
deduce $T_{\rm pt}$ from Eq.~(\ref{tpt}). 
We present our numerical results for $T_{\rm pt}$ in Fig.~\ref{fig11}
where it is represented parametrically 
as a function of the ground state energy, 
$E_{\rm gs} \propto (\phi - \phi_J)^2$.
For comparison, we also represent the crossover scale $T^\star$
directly deduced from the study of the density of states $d(\omega)$,
as discussed above.
 
The numerical results indicate that $T_{\rm pt}$ is 
larger than $T^{\star}$ at all investigated volume fractions 
by several orders of magnitude, 
typically more than a factor $10^4$. 
Also, the density dependence of $T_{\rm pt}$ is quite different from the
one of $T^\star$, since the numerical data seem to indicate 
that $T_{\rm pt} \sim (E_{\rm gs}/N)^{1/2} \sim (\phi-\phi_J)$.  
This means that the harmonic-anharmonic transition observed in the 
DOS {\it cannot} be explained through a 
perturbative analysis around the ground state. These results
show that thermally excited packings near jamming are {\it not} 
stabilized by quartic terms in the expansion of the energy,
and another explanation has therefore to be sought for the existence 
of $T^\star$. 

\subsection{`Fragility' of contact network and 
non-analyticity of pair potential}

The above calculation is perturbative in nature. It amounts to assuming 
that when heated, the system will simply explore the energy minimum 
used to compute the dynamical matrix. Crucially, this
reasoning implicitly assumes 
that the forces are analytic functions, which is not correct for the
harmonic pair potential we use which is not analytic in $r=\sigma$ 
where it is truncated. In fact, the above calculation 
would hold if $v(r) = (1-r/\sigma)^2$ for all $r$-values. Indeed, 
it was found in Ref.~\cite{ohern} that the threshold for anharmonicity
for the non-truncated potential is about $10^4$ larger than for the truncated
potential. This numerical finding is in fact 
in quantitative agreement with the results of the previous 
section, recall the data shown in Fig.~\ref{fig11}. 

Therefore, we follow Ref.~\cite{ohern} and 
explore the idea that the change in the DOS observed 
above $T^\star$ is associated 
with the breaking and renewal of the `contact network'. 
In such case, the anharmonicity would directly result from the non-analyticity 
in the pair potential. More precisely, it could be that 
some contacts are lost (or created) when $T$ increases 
long before the high-order non-linearities become
relevant. This physics cannot be 
described within a perturbative expansion 
around the ground state and is thus nonperturbative in nature.

In order to get a quantitative estimate of $T^{\star}$ we 
must determine the  temperature
scale for which thermal fluctuations become large enough to 
disrupt the contact network. We start again 
with the jammed phase. 
Our physical idea is that since overlaps between particles are 
small close to jamming, they are more rapidly blurred by the 
effect of thermal vibrations. 

Let us now make this idea more precise. 
At volume fraction $\phi > \phi_J$ and $T=0$, 
the typical length of the overlap 
between  contacting particles is 
$\delta_0 = \sigma - |{\bf r}_i - {\bf r}_j| \approx (\phi-\phi_J)$.
The latter estimate comes from the scaling of the width of  
the first peak in the pair correlation function near $\phi_J$. 
On the other hand, at finite temperature, the overlap between 
two particles becomes a fluctuating quantity. 
The typical displacement between two neighboring particles
has different scaling depending on the direction, longitudinal (L) or 
transverse (T), to the bond vector ${\bf r}_i - {\bf r}_j$~\cite{hecke}:
\be
\delta_L\propto \sqrt{T}, \qquad \delta_T\propto 
\sqrt{T} \, (\phi-\phi_J)^{-1/4} \,.
\ee 
In order to loose a contact one must have either $\delta_L \sim \delta_0$ or 
$\delta_T \sim \sqrt{\delta_0}$, from purely geometrical reasoning. The 
first identity is more constraining, 
and implies that the notion of `contacts' between
particles only makes sense when $\sqrt{T} < (\phi - \phi_J)$, which 
directly explains the scaling of the crossover 
temperature $T^\star \sim (\phi-\phi_J)^2$.

On the hard sphere side below jamming, a similar reasoning holds with the
overlap being replaced by the gap between the particles. 
Note that this argument also implies that the quadratic dependence 
of $T^\star$ with the distance to $\phi_J$ is actually
a direct consequence of the exponent in the pairwise 
interaction potential, which we have considered to be purely 
harmonic throughout this study.

We can get an even more quantitative grasp of the crossover scale 
$T^\star$ through an analysis of the radial distribution function $g(r)$.
At $T=0$, $g(r)$ has a sharp peak near $r=\sigma$, 
and its width is $\delta_0 \simeq  0.4 (\phi-\phi_J)$, the prefactor
being determined numerically.
In general, the radial distribution function is a volume 
average of the radial distribution function around a specific particle, 
\be
g(r) = \frac{1}{N} \sum_i 
g_{i}(r) = \frac{1}{N} \sum_i \left\langle \sum_j \delta(r-|{\bf r}_{ij}|) 
\right\rangle. 
\ee 
At $T=0$, $g_{i}(r)$ is the sum of delta functions corresponding to 
the relative distance between particle $i$ and all its neighbors
in the studied packing. When $T > 0$, 
these delta functions broaden under the influence of thermal 
fluctuations, and thus they acquire a
width of the order $\sqrt T$ discussed above. 

\begin{figure}
\psfig{file=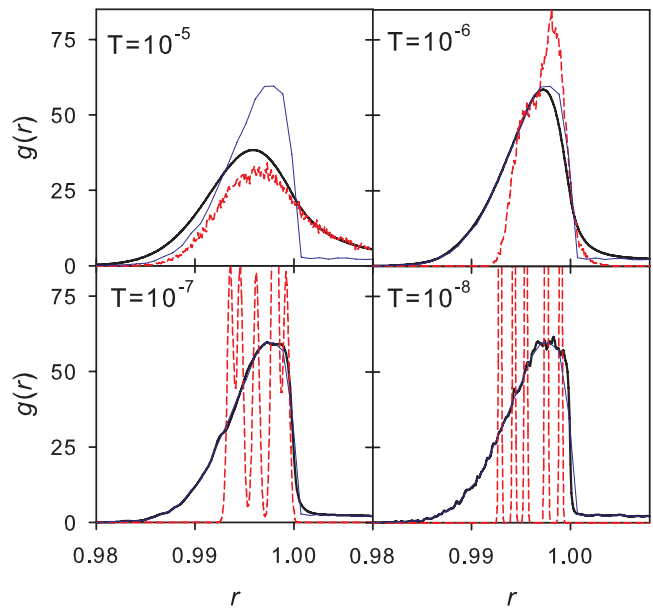,width=8.5cm,clip}
\caption{\label{fig9} 
The radial distribution function $g(r)$ at finite
$T$, (black / thick lines), at $T=0$ (blue / thin line), and its 
particle-resolved version $g_{i}(r)$ (red / dashed lines)
for a randomly chosen particle $i$ at $\phi=0.66$ and various temperatures. 
While contacts between $i$ and its neighbors are well-defined 
at low $T$, they cannot be resolved when $T$ becomes 
larger than $T^{\star}$ which is $\approx 4 \times 10^{-7}$ 
for this particular density.}
\end{figure}

In Fig.~\ref{fig9},
we presents numerical results for $g(r)$ and the $g_{i}(r)$ 
for a randomly chosen particle $i$ at various temperatures both above
and below $T^\star$, which for this example 
is $T^\star \approx 4 \cdot 10^{-7}$. 
At $T=10^{-8}$, $g_{i}(r)$ shows well-resolved 
delta functions, while $g(r)$ is close to its $T=0$ value.
With increasing temperature, the peaks in $g_{i}(r)$ broaden, 
and cannot be resolved  anymore above $T^\star$ when their width  
becomes comparable to the width of the $T=0$ width of $g(r)$. At even 
higher temperature, $T=10^{-5}$, the overall shape 
of $g(r)$ is now controlled by the broadening of individual 
$g_{i}(r)$, and its shape becomes very different 
from the structure at $T=0$, see Ref.~\cite{hugopre}. 

Let us push the argument further. 
Quantitatively, we determine numerically that the 
thermal broadening of individual
peaks in $g_i(r)$ is given by $\simeq 0.9 \sqrt{T}$, such 
that at $T=T^\star$ we obtain a width $\simeq 0.03 (\phi-\phi_J)$. 
This should be compared to the width of $g(r)$ at half maximum,
$\delta_0 \approx 0.4 (\phi-\phi_J)$, which is about 10 times
larger. This numerical factor is reasonable since each particle 
possesses on average about 6 neighbors, and so the smallest
of these overlaps is about $\delta_0 / 6 \sim 0.066 (\phi-\phi_J)$ 
which is indeed not far from $0.03 (\phi-\phi_J)$. 

Our argument suggests that anharmonicy starts to play a role
and modify the DOS when an {\it extensive} number of contacts 
become affected by thermal fluctuations. 
This is at odds with the results in Ref.~\cite{ohern}
where a threshold for anharmonicity was defined via the breaking 
of a {\it single} contact in a given packing. This 
qualitative difference in our analysis presumably explains 
the different conclusion that we reach in this work, namely that 
a finite regime exists where a harmonic approximation 
is valid even in the thermodynamic limit,
thus contradicting the different claim made in 
Ref.~\cite{ohern} that the domain of validity of harmonic 
theory is null.

To summarize, we find that the onset of anharmonicity 
does not follow from the breakdown of a perturbative 
expansion to second order, but rather stems from the 
non-analycity of the pair interaction. 
Remark that the use of a truncated pair potential is 
of course a mandatory ingredient for the system to display 
a jamming transition in the first place. There is thus
no room, it would seem, to make the physics associated to the 
jamming transition more robust against thermal fluctuations
since the very origin of the critical fluctuations---a 
marginally stable solid with vanishingly small overlaps 
between particles---also makes it highly sensitive 
to thermal fluctuations. This conclusion suggests that 
experimental investigations of the jamming transition 
using colloidal particles are not straightforward, 
as we now discuss. 

\section{Discussion: Experiments versus theory}

\label{discussion}

The theoretical analysis conducted in this article shows 
that single particle dynamics is able to directly reveal
the divergence of time scales and length scales when the 
jamming transition at ($T=0$, $\phi=\phi_J$) is approached. 
We have additionnally shown that dynamic criticality survives 
a finite amount of temperature, and have carefully explained 
under what conditions the signatures of the jamming singularity
remain observable in the dynamics at finite temperature, 
emphasizing in particular the existence of three distinct 
critical regimes in the vicinity of the critical point, 
Fig.~\ref{fig10a}. We suggested a possible connections 
between the vibrational heterogeneity unveiled here and the 
experimental reports in Ref.~\cite{leche,corentin}. Other
experiments report a nonmonotonic evolution of dynamic 
heterogeneity with density~\cite{luca1,luca2}, but the connection 
with our work is less clear.   
 
We mentioned in the introduction the important experimental 
activity in the colloidal community aiming at characterizing 
the emergence of `soft modes' near the jamming transition.  
It is interesting to use our findings to discuss these experimental 
investigations. Two different types of colloidal systems 
have been used, which we discuss separately. 

\begin{figure}
\psfig{file=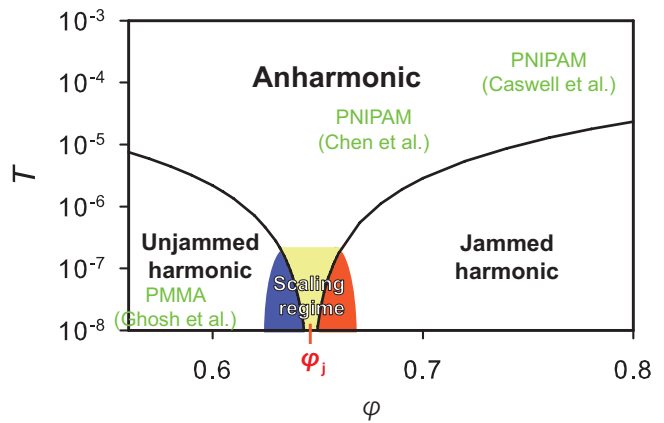,width=8.5cm,clip}
\caption{\label{fig10} An enlarged view of 
the (Temperature, Volume fraction) phase diagram 
reporting the critical regimes shown in Fig.~\ref{fig10a}, 
and the approximate location of the experimental
studies aimed at studying anomalous vibrational
dynamics in colloidal systems: 
Gosh {\it et al.}~\cite{bonn} use PMMA hard spheres,
while Chen {\it et al.}~\cite{soft} and
Caswell {\it et al.}~\cite{caswell} study PNIPAM microgel 
particles.  Previous studies all  
lie too far away from the jamming transition 
to detect the dynamic criticality associated to the transition.}
\end{figure}

A first series of experiments concentrates on colloidal 
PMMA hard spheres~\cite{bonn,bonn2,xuprl}.
In that case, temperature simply serves to produce Brownian motion 
but since particles are `infinitely' hard, temperature does 
not blur the jamming transition, and thus these experiments 
are in principle well-suited to probe Regime II, the critical 
regime of hard spheres. However, the investigations in 
Refs.~\cite{bonn,bonn2} were performed in 
the range $\phi=0.56 - 0.60$. Although some care 
must be taken with the absolute values of $\phi$ in experimental 
work, these values seem consistent because the 
amplitude of the experimentally-measured DW factors are 
in the range $\Delta^2(\infty) \approx (0.005 - 0.03)\sigma^2$, 
see Fig.~\ref{fig1}. However, these volume fractions 
are very far from $\phi_J$ (see Figs.~\ref{fig4} and 
\ref{fig5}) and all lie outside the critical regime II
where a sharp drop in the DW factor would be observed, accompanied 
by the appearance of low-frequency modes and a growing correlation length. 
It would therefore be interesting to push these experiments further 
towards the jamming transition, to see whether 
experimental signatures of the dynamic criticality discussed 
in this work can be detected.
In Fig.~\ref{fig10}, we provide an extended view of the 
($T, \varphi$) phase diagram including the experimentally explored 
regime for hard spheres, outside the critical region discussed 
in our work.

A second series of experiments designed to probe low-frequency modes 
near jamming uses soft `PNIPAM'
microgel particles~\cite{soft,craig,craig2,caswell}. 
In this case both the volume fraction and the temperature, 
expressed in units of the particle softness, are relevant control parameters. 
Because microgel particles are easily compressed, 
the volume fraction can be adjusted to be as close 
to the jamming density as desired. However, softness 
is now the main issue. The adimensional 
temperature estimate is 
$T/\epsilon \approx 2 \times 10^{-5}$ in Ref.~\cite{soft}. We can 
confirm this value by again using the DW factor as a sensitive `thermometer'. 
Indeed, the values measured near jamming, $\Delta^2(\infty) 
\sim 3 \times 10^{-4} \sigma^2$, correspond in our Fig.~\ref{fig1} 
to a temperature scale of about $10^{-5}$. In very recent work 
with different microgel particles~\cite{caswell},
the DW factor is larger, $\Delta^2(\infty) 
\sim 2 \times 10^{-3} \sigma^2$, 
and thus the system is effectively at an even higher 
temperature, of the order $10^{-4}$.
On the basis of our numerical results, we conclude that the
microgel particles studied in Refs.~\cite{soft,caswell} are so soft 
(or, equivalently so `hot')
that anharmonicity is very strong because 
temperature is much larger than the crossover scale $T^\star$, 
see their location 
in the phase diagram of Fig.~\ref{fig10}. As for hard colloids, we suggest that
it could be interesting to design different experimental systems 
with particles that are less soft than microgels, for instance
emulsions,  in order to probe experimentally the critical dynamics
associated to the jamming transition. 

In conclusion, because of the smallness of the 
critical region (that certainly would deserve a theoretical explanation) 
experiments have not yet probed critical 
properties related to the jamming transition. Since they
are quite far from the critical region, our results imply that  
there cannot be any direct connection between the DOS 
measured in those experiments
and the theoretical results established at $T=0$.   
On the basis of our results, one should find an alternative 
explanation, not based 
on `thermal vestiges' of the jamming transition, for the experimental results.  
Indeed, there is no need to invoke any critical behavior 
to account for the smooth evolution with density  
of the low-frequency $\omega^\star$ reported in Ref.~\cite{soft}.
It is fully consistent with the data obtained in Fig.~\ref{fig2}
for large temperatures which basically track the microscopic 
time scale $\tau_0$ that changes smoothly across $\phi_J$ 
as a direct result of the compression.  
Another experimental signature of the jamming singularity
is the nonmonotonic evolution of the 
pair correlation upon compression~\cite{caswell,zhang}.
This can also be explained in a way unrelated to the critical 
properties of the jamming 
transition since, as demonstrated in Ref.~\cite{hugosoft}, 
this anomaly persists arbitrarily far 
from the transition, even in the equilibrium fluid.

In conclusion, our study reveals that single particle motion
in thermal systems near jamming becomes singular, 
and reveals a number of scaling laws and divergences
for both time scales and length scales. This dynamic criticality
thus provides direct evidence of the critical nature 
of the jamming transition occurring at $T=0$, and 
might suggest new routes to attack the problem from a theoretical 
viewpoint, for instance using the replica approach in the 
framework of statistical mechanics. In addition, we have suggested 
that it would be very interesting to design 
new experiments using colloidal systems to probe more
closely the diverging time scales and length scales 
associated to vibrational dynamics of dense packings.  

\acknowledgments

We thank O. Dauchot, C. Goodrich, and F. Zamponi for useful 
discussions. Our work is supported by R\'egion Languedoc-Roussillon 
(AI, LB), and from ERC grant NPRG-GLASS (GB).

\appendix

\section{Derivation of Eqs.~(\ref{ptg} - \ref{pt2})}

Using the normal modes decomposition of the position vector of a 
particle Eq.~(\ref{rex}), 
the potential energy of the system can be expanded as 
\begin{eqnarray}
E &=& E_{\rm gs} + \frac{1}{2} \sum_{ab} V_{ab} x_a x_b
+ \frac{1}{3!} \sum_{abc} V_{abc} x_a x_b x_c \nonumber \\
&& + \frac{1}{4!} \sum_{abcd} V_{abcd} x_a x_b x_c x_d + 
\mathcal{O}(x^5), \label{apot}
\end{eqnarray}
where we introduced a simplified notation for the derivatives, e.g. 
$V_{ab} = \frac{\partial^2 v}{\partial x_a \partial x_b}$. 
Note that the first order term vanishes since we focus on a  
potential energy minimum. 
We regard this development of the potential energy as the sum of the 
harmonic term $E_0$ (up to second order) 
and anharmonic terms $\Delta E$. 
When $\Delta E$ is treated as perturbation, we can obtain the following 
expansion of the distribution function, 
\begin{eqnarray}
f &=& f_0 \Bigl( 1 - \beta (\Delta E - \ave{\Delta E}_0) + 1/2 \beta^2 
(\Delta E^2 - \ave{\Delta E^2}_0 \nonumber \\
&& - 2\Delta E\ave{\Delta E}_0 + 2\ave{\Delta E}_0^2) + \mathcal{O}(\beta^3 
\Delta E^3) \Bigr),  \label{adis}
\end{eqnarray}
where $f$ is the canonical distribution function corresponding to $E$ and 
$f_0$ to $E_0$;  
$\ave{ \cdots}_0$ indicates a thermal average over the distribution $f_0$.  
Taking the thermal average of Eq.~(\ref{apot}) using 
the distribution function in Eq.~(\ref{adis}), 
we obtain the perturbation expansion of the potential energy.
 
To organize this series as the low-temperature expansion, 
we have estimated the temperature dependence of each term using the 
fact that $\ave{x^n}_0 \propto T^{n/2}$ when $n$ is even.
As a result, we get the following equation:
\begin{widetext}
\begin{eqnarray}
\ave{E} &=& E_{gs} + \frac{1}{2} \sum_{ab} V_{ab} \ave{x_a x_b}_0 + 
\frac{1}{4!} \sum_{abcd} V_{abcd} \ave{x_a x_b x_c x_d}_0 
- \frac{1}{T} \sum_{abcdef} \Bigl( \frac{1}{3!3!} V_{abc} V_{def} 
\ave{x_a x_b x_c x_d x_e x_f}_0 \nonumber \\ 
&& + \frac{1}{2!4!} V_{ab} V_{cdef} \ave{x_a x_b x_c x_d x_e x_f}_0 
- \frac{1}{2!4!}  V_{ab} V_{cdef}\ave{x_a x_b}_0 \ave{x_c x_d x_d x_f}_0)  
\Bigr) \nonumber \\
&& + \frac{1}{2T^2} \sum_{abcdefgh} \frac{1}{2!3!3!} V_{ab} V_{cde} 
V_{fgh} \Bigl( \ave{x_a x_b x_c x_d x_e x_f x_g x_h}_0 - 
\ave{x_a x_b}_0\ave{x_c x_d x_e x_f x_g x_h}_0 \Bigr) + 
\mathcal{O}(T^3),  
\end{eqnarray}
\end{widetext}
where the first term is constant, the second term is proportional to $T$, 
the next 6 terms are ${\cal O}(T^2)$, and the rest is $\mathcal{O}(T^3)$.  
Although four-, six- and eight-point functions appear in this 
expression, they can be evaluated using the Gaussian 
approximation and Wick's theorem. 
For example, the expectation value of the two-point function is simply
\be
\ave{x_a x_b}_0 = \frac{\delta_{ab} T}{\lambda_a}, 
\ee 
where $\lambda_a$ is 
the eigenvalue for the $a$-th normal modes. 
Note also that the second order derivatives are the diagonal matrix $V_{ab} 
= \lambda_a \delta_{ab}$. 
Using these relations with Wick's theorem, we obtain the final expressions 
in Eqs.~(\ref{ptg}-\ref{pt2}).


\begin{thebibliography}{99}

\bibitem{reviewmodes}
A. J. Liu, M. Wyart, W. van Saarloos and S. R. Nagel
in {\it Dynamical heterogeneities in glasses, 
colloids and granular materials}, 
Eds.: L. Berthier, G. Biroli, J.-P. Bouchaud,
L. Cipelletti, and W. van Saarloos, (Oxford University Press, Oxford, 2011).

\bibitem{wyart2}
M. Wyart, L. Silbert, S. R. Nagel, and T. A. Witten,
Phys. Rev. E {\bf 72}, 051306 (2005).

\bibitem{silbert}
L. E. Silbert, A. J. Liu, and S. R. Nagel, 
Phys. Rev. Lett. {\bf 95}, 098301 (2005).

\bibitem{xu}
L. Wang and N. Xu, arXiv:1112.2429.

\bibitem{vitelli}
N. Xu, V. Vitelli, A. J. Liu and S. R. Nagel,
Europhys. Lett. {\bf 90}, 56001 (2010).

\bibitem{vitelli2}
V. Vitelli, N. Xu, M. Wyart, A. J. Liu, and S. R. Nagel, 
Phys. Rev. E {\bf 81}, 021301 (2010).

\bibitem{recentwyart}
G. D\"uring, E. Lerner, and M. Wyart,
arXiv:1204.3542.

\bibitem{brito}
C. Brito and M. Wyart, J. Chem. Phys. {\bf 131}, 024504 (2009).

\bibitem{bonn}
A. Ghosh, V. K. Chikkadi, P. Schall, J. Kurchan, and D. Bonn,
Phys. Rev. Lett. {\bf 104}, 248305 (2010).

\bibitem{bonn2}
A. Ghosh, R. Mari, V. Chikkadi, P. Schall, J. Kurchan, and D. Bonn, 
Soft matter {\bf 6}, 3082 (2010). 

\bibitem{soft}
K. Chen, W. G. Ellenbroek, Z. Zhang, D. T. N. Chen, P. J. Yunker, 
C. Brito, O. Dauchot, S. Henkes, W. van Saarloos, A. J. Liu, and A. G. Yodh, 
Phys. Rev. Lett. {\bf 105}, 025501 (2010). 

\bibitem{craig}
D. Kaya, N. Green, C. E. Maloney and M. F. Islam. 
Science {\bf 329}, 656 (2010). 

\bibitem{craig2}
D. Kaya, N. Green, C. E. Maloney and M. F. Islam. 
Phys. Rev. E {\bf 83}, 051404 (2011).

\bibitem{xuprl}
P. Tan, N. Xu, A. B. Schofield, and L. Xu, 
Phys. Rev. Lett. {\bf 108}, 095501 (2012).

\bibitem{caswell}
T. A. Caswell, Z. Zhang, M. L. Gardel, S. R. Nagel, 
arXiv:1206.6802 (2012).

\bibitem{britosoft} 
C. Brito, O. Dauchot, G. Biroli, and J.-P. Bouchaud,
Soft Matter {\bf 6}, 3013 (2010).

\bibitem{chaikin}
P. M. Chaikin and T. C. Lubensky,
{\it Principles of Condensed Matter Physics},
(Cambridge University Press, Cambridge, 1995).

\bibitem{tom}
L. Berthier and T. A. Witten,  EPL {\bf 86}, 10001 (2009); 
Phys. Rev. E {\bf 80}, 021502 (2009).

\bibitem{hugosoft}
H. Jacquin and L. Berthier, Soft Matter {\bf 6}, 2970 (2010). 

\bibitem{hugopre}
L. Berthier, H. Jacquin, and F. Zamponi,
Phys. Rev. E {\bf 84}, 051103 (2011); 
J. Stat. Mech. P01004 (2011).

\bibitem{hayakawa}
M. Otsuki and H. Hayakawa, arXiv:1111.1313

\bibitem{ohern}
C. F. Schreck, T. Bertrand, C. S. O'Hern, and M. D. Shattuck, 
Phys. Rev. Lett. {\bf 107}, 078301 (2011).

\bibitem{dauchotmodes}
S. Henkes, C. Brito and O. Dauchot, Soft Matter {\bf 8}, 6092 (2012).

\bibitem{PZ} G. Parisi and F. Zamponi, Rev. Mod. Phys. {\bf 82}, 789 (2010).

\bibitem{hugo}
H. Jacquin, L. Berthier, and F. Zamponi,
Phys. Rev. Lett. {\bf 106}, 135702 (2011).

\bibitem{leche}
F. Lechenault, O. Dauchot, G. Biroli, and J.-P. Bouchaud, 
Europhys. Lett. {\bf 83}, 46002 (2008).

\bibitem{corentin}
C. Coulais, R. P. Behringer, and O. Dauchot,
arXiv:1202.5687.

\bibitem{saarloos}
W. G. Ellenbroek, E. Somfai, M. van Hecke, W. van Saarloos,
Phys. Rev. Lett. {\bf 97}, 258001 (2006). 

\bibitem{durian}
D. J. Durian, Phys. Rev. Lett. {\bf 75}, 4780 (1995).

\bibitem{ohern1}
C. S. O'Hern, S. A. Langer, A. J. Liu, and S. R. Nagel,
Phys. Rev. Lett. {\bf 88}, 075507 (2002).

\bibitem{pinaki}
P. Chaudhuri, L. Berthier, and S. Sastry,
Phys. Rev. Lett. {\bf 104}, 165701 (2010).

\bibitem{teitel}
P. Olsson and S. Teitel, Phys. Rev. Lett. {\bf 99}, 178001 (2007).

\bibitem{claus}
C. Heussinger, L. Berthier, and J.-L. Barrat,
EPL {\bf 90}, 20005 (2010).

\bibitem{ikeda}
A. Ikeda, L. Berthier, and P. Sollich,
Phys. Rev. Lett. {\bf 109}, 018301 (2012).

\bibitem{rmp} L. Berthier and G. Biroli, 
Rev. Mod. Phys. {\bf 83}, 587 (2011). 

\bibitem{oldtom}
G. S. Grest, S. R. Nagel, A. Rahman, and T. A. Witten, 
J. Chem. Phys. {\bf 74}, 3532 (1981).

\bibitem{hansen}
J. P. Hansen and I. R. McDonald, {\it Theory of Simple Liquids},
(Elsevier, Amsterdam, 1986).

\bibitem{reviewchi4}
L. Berthier, G. Biroli, J.-P. Bouchaud, and R. L. Jack, 
in {\it Dynamical heterogeneities in glasses, colloids, and 
granular media}, 
Eds.: L. Berthier, G. Biroli, J.-P. Bouchaud,
L. Cipelletti, and W. van Saarloos, (Oxford University Press, Oxford, 2011).

\bibitem{cristina}
C. Toninelli, M. Wyart, L. Berthier, G. Biroli, and J.-P. Bouchaud,
Phys. Rev. E {\bf 71}, 041505 (2005).

\bibitem{jcp}
L. Berthier, G. Biroli, J.-P. Bouchaud, W. Kob, K. Miyazaki, and
D. Reichman, J. Chem. Phys. {\bf 126}, 184503 (2007);
J. Chem. Phys. {\bf 126}, 184504 (2007).

\bibitem{rmp77}
P. C. Hohenberg and B. I. Halperin,
Rev. Mod. Phys. {\bf 49}, 435 (1977).

\bibitem{hatano}
T. Hatano, Phys. Rev. E {\bf 79}, 050301(R) (2009).

\bibitem{drocco}
J. A. Drocco, M. B. Hastings, C. J. Olson Reichhardt,
and C. Reichhardt, Phys.Rev. Lett. {\bf 95}, 088001 (2005).

\bibitem{jorge}
J. Kurchan, G. Parisi, and F. Zamponi, arXiv:1208.0421

\bibitem{martin}
L. R. G\'omez, A. M. Turner, M. van Hecke, and V. Vitelli,
Phys. Rev. Lett. {\bf 108}, 058001 (2012).

\bibitem{hecke}
W.G. Ellenbroek, E. Somfai, M. van Hecke, W. van Saarloos, 
Phys. Rev. Lett. {\bf 97} 258001 (2006).

\bibitem{zhang}
Z. Zhang, N. Xu, D. T. N. Chen, P. Yunker, A. M. Alsayed,
K. B. Aptowicz, P. Habdas, A. J. Liu, S. R. Nagel,
and A. G. Yodh, Nature {\bf 459}, 230 (2009).

\bibitem{luca1}
P. Ballesta, A. Duri, and L. Cipelletti, 
Nature Phys. {\bf 4}, 550 (2008).

\bibitem{luca2}
D. A. Sessoms, I. Bischofberger, L. Cipelletti, and V. Trappe,
Phil. Trans. R. Soc. A {\bf 367}, 5013 (2009).

\end{thebibliography}
\end{document}